\documentclass[12pt]{article}

\usepackage[margin=0.75in]{geometry}
\usepackage{amsmath,amssymb,amsthm,mathtools, bm}
\usepackage{comment}
\usepackage{booktabs}
\usepackage{float}
\usepackage{hyperref}
\usepackage{enumitem}
\usepackage{tikz}
\usepackage{algorithm}
\usepackage{algpseudocode}
\usetikzlibrary{arrows.meta, calc, positioning, decorations.pathreplacing}

\newtheorem{theorem}{Theorem}
\newtheorem{lemma}{Lemma}

\newtheorem{example}{Example}
\newtheorem*{example*}{Example}
\newcommand{\ket}[1]{\lvert #1\rangle}
\newcommand{\bra}[1]{\langle #1\rvert}

\newcommand{\cC}{\mathbb{C}}
\newcommand{\cF}{\mathbb{F}}

\title{Coherent Recovery of Nested Codes over the Quantum Erasure Channel Using Erasure Decoding and Quantization}

\author{
  \\ Aditya Guhagarkar,  \quad S. Sandeep Pradhan, and \quad Achilleas Anastasopoulos \\
  \small Department of Electrical and Computer Engineering \\
  \small University of Michigan, Ann Arbor \\
  \small email: \{adityg, pradhanv, anastas\}@umich.edu
}
\date{\today}

\begin{document}

\maketitle

\begin{abstract}
Motivated by quantum capacity results, we revisit nested binary linear codes for the quantum erasure channel through a packing--covering viewpoint: the outer code supports erasure reconstruction, while the inner code performs erasure quantization. We turn this perspective into two explicit decoder architectures. Architecture~I requires no stabilizer-syndrome measurements and combines a classical peeling decoder and erasure quantizer with corresponding coherent quantum reconstruction and cleanup operations that disentangle the logical state from inaccessible erasure-environment branches. Architecture~II goes one step further and implements erasure belief propagation directly through reversible ternary quantum gates, providing quantum state healing that reconstructs erased symbols while leaving unerased code states unperturbed. We establish sufficient conditions for exact recovery of arbitrary logical states, and illustrate both architectures using the Steane code, including explicit scheduling procedures and coherent quantum circuits.
\end{abstract}

\clearpage

\section{Introduction}
\label{sec:introduction}

The quantum erasure channel (QEC) models the loss of quantum information when the receiver can identify the affected physical systems, as in photon loss with known missing-photon locations \cite{bennett1997capacities,grassl1997codes}. Each transmitted qubit either arrives unchanged or is replaced by an orthogonal erasure flag. 
Knowledge of the erased positions aids recovery but does not reveal the quantum information they contained. The erased qubits dissipate into the environment, implying that they will be partially traced over. In other words, given a fixed erasure pattern, the pure quantum encoded state unravels into a mixture state. A successful receiver must recover arbitrary logical states while preserving their coherence.

The unassisted quantum capacity of the erasure channel with erasure probability $\epsilon$ is $\max\{0,1-2\epsilon\}$ qubits per channel use \cite{bennett1997capacities}. The connection between private classical communication and quantum-capacity achievability motivates a packing--covering viewpoint: the receiver must reconstruct the transmitted information, while the environment must learn asymptotically nothing about the logical input~\cite{Devetak2005}. For nested binary linear codes $\cC_2\subseteq\cC_1\subseteq\mathbb F_2^n$, the outer code $\cC_1$ supports reconstruction from surviving coordinates, while superposition over the inner code $\cC_2$ supplies the scrambling or  randomization needed to conceal logical information from the erased subsystem. The logical basis states are uniform superpositions over cosets of $\cC_2$ in $\cC_1$, giving quantum rate $R_Q=R_1-R_2$, where $R_i=\dim(\cC_i)/n$.
For erasures, the covering task can be expressed through binary erasure quantization: finding a codeword that agrees with prescribed binary values on selected coordinates, while leaving the remaining coordinates unconstrained \cite{martinian2003iterative}. This packing--covering viewpoint motivates nested code families with rates approaching $R_1=1-\epsilon$ and $R_2=\epsilon$, yielding a quantum rate $R_Q=R_1-R_2$ approaching $1-2\epsilon$.

Calderbank--Shor--Steane (CSS) codes provide a natural setting for this problem because their construction from nested binary linear codes connects quantum recovery with classical coding methods~\cite{CalderbankShor1996,steane1996simple}.
The conventional CSS description emphasizes the channel codes $\cC_1$ and $\cC_2^\perp$, associated with complementary error types. This viewpoint has supported capacity-achieving constructions, including CSS codes based on binary Reed--Muller codes \cite{kumar2016reedmuller}. Martinian and Yedidia relate erasure decoding in $\cC_2^\perp$ to erasure quantization in $\cC_2$ through feasibility and algorithmic duality \cite{martinian2003iterative}. Our packing--covering formulation emphasizes how the latter task enters the receiver: a decoding guarantee for $\cC_2^\perp$ alone does not specify the coherent completion operation required in $\cC_2$. This particular duality is specific to the erasure problems and does not establish a general equivalence between good packing and covering codes for arbitrary channels or distortion measures.
We depart from the CSS viewpoint, and 
espouse the quantum capacity-inspired packing-covering codes perspective, with erasure recovery based on $\mathbb{C}_1$ and erasure quantization based on $\mathbb{C}_2$.

In conventional stabilizer-based erasure recovery, measured syndromes and known erasure locations guide a classical decoder in determining a Pauli correction \cite{Gottesman1997}. For quantum low-density parity-check codes, graphical methods include BP with guided decimation \cite{gokduman2024erasure}, degeneracy-aware BP \cite{kuo2026degenerate}, and peeling combined with cluster decomposition and Gaussian elimination \cite{yao2025cluster}. Recovery without explicit syndrome measurement also has established precedents. Li et al.\ describe unitary recovery followed by a partial trace \cite{li2012recovery}; Matsumoto and Chiwaki--Matsumoto develop unitary and measurement-free reconstruction for quantum secret sharing \cite{matsumoto2017unitary,chiwaki2025measurementfree}. Nakata, Matsuura, and Koashi further connect classical decoding with quantum recovery for CSS and more general quantum codes \cite{nakata2025decoding}.

We give an explicit coherent erasure-recovery construction in which classical reconstruction and erasure quantization have complementary operational roles. Tracing out erased qubits generally leaves a mixture of branches whose labels are inaccessible to the receiver. Coherent reconstruction restores valid computational-basis codewords within each branch, but does not generally restore the original encoded pure state. Inverse encoding exposes the logical register while leaving a logical-label-dependent constraint on the auxiliary register. The encoder coordinates of a suitable completion in $\cC_2$ determine an auxiliary shift that removes this dependence. Evaluating this shift coherently completes logical recovery without access to the erased values or branch labels. Together, these stages provide a form of quantum state healing: they restore the logical quantum information while separating it from the residual degrees of freedom associated with erasure.

Our contributions are as follows:
\begin{enumerate}
    \item \textbf{Classically scheduled coherent recovery.} Architecture~I combines a peeling-reconstruction schedule for $\cC_1$ with a quantization reservation schedule for $\cC_2$. Both schedules depend only on the code graphs and the observed erasure pattern, without evaluating unknown quantum-register values. We implement the scheduled binary operations coherently using temporary workspace cleared by reversing the computation. The proposed decoder does not perform any syndrome measurement.

    \item \textbf{Sufficient conditions for exact logical recovery.} We show how reconstruction, inverse encoding, and quantization-based cleanup yield
    \begin{equation}
\rho_{\mathrm{out},E}=\ket{\psi_L}\!\bra{\psi_L}\otimes\sigma_E
    \end{equation}
    when reconstruction succeeds, and the quantizer supplies valid completions for the required targets. The residual state $\sigma_E$ is independent of the logical input $\ket{\psi_L}$, and all temporary computational workspace returns to zero.

    \item \textbf{Quantum BP reconstruction.} Architecture~II performs reconstruction using reversible ternary check and variable updates on quantum registers, without a classically prepared peeling schedule for that stage. It retains the inverse-encoding and quantization-based cleanup stages of Architecture~I. The comparison concerns the implementation of reconstruction and does not presume a lower overall cost for quantum BP.
\end{enumerate}

The emphasis of this work is to turn the packing--covering description of quantum coding into an explicit recovery procedure. We show how coherent reconstruction, inverse encoding, and quantization-based cleanup work together to recover the complete logical state, and establish sufficient conditions for their success. This provides a framework for implementing recovery with capacity-achieving nested code families: the constituent reconstruction and quantization procedures must succeed with vanishing joint failure probability.

The proposed quantum decoder can be thought of as a quantum entropy extractor. Assuming that a pure logical quantum state is encoded and sent over the channel,
at the erasure channel output, the state has von Neumann entropy of approximately $2n\epsilon$ qubits, distributed across $n$ coded qutrit states. 
There is $n\epsilon$ qubits of entropy in the erasure pattern, and  for every typical erasure pattern, there is
$n \epsilon$ entropy of environment branches.
After the proposed quantum state healing, the entropy will be distilled into  $2n\epsilon$ bits/qubits.
The quantum channel decoder will extract $n \epsilon$ bits of entropy, and the quantum erasure quantizer will extract $n \epsilon$ bits of entropy. At the end,  $2n \epsilon$ bits/qubits of ancillas having random states, will need to be dissipated into the environment. 
The healing will leave the state unperturbed if there are no erasures. 
This approach could be related to Maxwell's demon \cite{maxwell2012theory,bennett1982thermodynamics}. The decoder is like a demon that has a nose for entropy, and extracts it to the ancillas.

The remaining of this paper is structured as follows.
Section~\ref{sec:preliminaries} introduces the nested-code construction, quantum erasure model, and branch representation. Section~\ref{sec:classical-architecture} develops Architecture~I and its recovery conditions. Section~\ref{sec:quantum-bp-architecture} presents Architecture~II and its integration with the shared inverse-encoding and cleanup stages.
We conclude in Section~\ref{sec:conclusions} by providing some directions for future 
research.

\section{Preliminaries}
\label{sec:preliminaries}


\subsection{QEC}

Let $\mathcal H_Q=\operatorname{span}\{\ket0,\ket1\}$ denote the Hilbert space of a qubit, and let $\ket e$ be an erasure state orthogonal to $\mathcal H_Q$. The QEC with erasure probability $\epsilon$ is defined as the following CPTP (completely positive trace-preserving) map
\begin{equation}
\label{eq:qec-channel}
    \mathcal N_\epsilon(\rho)
    =
    (1-\epsilon)\rho
    +
    \epsilon\ket e\bra e.
\end{equation}
The environment sees the following complementary channel
\[
\mathcal{N}^C_{\epsilon}(\rho) = \epsilon \rho+ (1-\epsilon) \ket{e}\bra{e}.
\]

The channel output belongs to the qutrit space
$\operatorname{span}\{\ket0,\ket1,\ket e\}$.
Since $\ket e$ is orthogonal to the computational subspace, the receiver can identify erasure locations by measuring with projectors:
\begin{equation}
    P_{\mathrm{valid}}
    =
    \ket0\bra0+\ket1\bra1,
    \qquad
    P_e=\ket e\bra e.
\end{equation}
These measurements identify the erased physical positions without revealing the computational-basis values of the surviving qubits.

The legitimate receiver sees 
$(1-\epsilon)$ fraction of qubits transmitted and sees erasure states for the rest. 
The environment sees $\epsilon$ fraction of the qubits (erased from the receiver), and see erasures for the rest. This implies that we have to partial trace over the erased qubits.
The encoder must ensure that the $\epsilon$ fraction of qubits that the environment sees provides no information about the message qubits. 
The quantum capacity of the erasure channel is 
\begin{equation}
\label{eq:qec-capacity}
    Q(\epsilon)=\max\{0,1-2\epsilon\},
\end{equation}
as established in~\cite{bennett1997capacities}.
The converse follows from the no-cloning theorem \cite{bennett1997capacities}. 
The achievability of the single-letter capacity formula uses random Hash coding, similar to random stabilizer codes.

Alternatively, a single-letter lower bound on the quantum capacity of memoryless general quantum channels was derived in 
\cite{shor2002quantum,Devetak2005}. 
For the QEC, the achievabilty uses the maximally mixed input state $\rho=\frac{1}{2} \ket{0}\bra{0}+\frac{1}{2}\ket{1}\bra{1}$. 
The coding strategy is inspired from the coding scheme for classical capacity of wiretap channels.  
The capacity is given by the coherent information 
\[
I_c(\rho,\mathcal{N}_{\epsilon})= S(\mathcal{N}_\epsilon(\rho))-S(\mathcal{N}_{\epsilon}^C(\rho)),
\]
where $S(\cdot)$ denotes the von Neumann entropy.
The coding scheme uses a nested coding scheme with a covering code $\mathbb{C}_2$ nested inside a packing code $\mathbb{C}_1$. The rates of the packing and covering codes, given by $R_1$ and $R_2$,  are equal to the Holevo quantum information \cite{holevo1973bounds} of the quantum channel and its complementary channel, respectively.  We have
\[
I_c(\rho,\mathcal{N}_{\epsilon})=R_1-R_2,
\]
\[
R_1=S(\mathcal{N}_\epsilon(\rho))-S(\mathcal{N}_\epsilon(\rho)|X)=1-\epsilon, \quad 
R_2=S(\mathcal{N}^C_\epsilon(\rho))-S(\mathcal{N}^C_\epsilon(\rho)|X)=\epsilon,
\]
where  $X$ is a classical-quantum random variable  taking values $\ket{0}$ and $\ket{1}$ with probability $1/2$ and $1/2$, and 
\[
S(\mathcal{N}^C_\epsilon(\rho)|X)=S(\mathcal{N}_\epsilon(\rho)|X)=\frac{1}{2}
S(\mathcal{N}_\epsilon(\ket{0}\bra{0}))+
\frac{1}{2}S(\mathcal{N}_\epsilon(\ket{1}\bra{1})).
\]
The packing code is for correcting the erasures incurred in the quantum channel. The covering code is a scrambling code that prevents leakage of quantum information to the environment. It can be interpreted as a code for covering a ternary source \cite{martinian2003iterative} as described in the sequel.

\paragraph{Binary erasure quantization problem:}  
This classical quantization problem was first considered in \cite{martinian2003iterative}, where a memoryless source produces one of three symbols $0$, $1$ and $e$, where $e$ denotes the erasure symbol, and  the probabilities of these symbols are given by $P_S(0)=P_S(1)=\epsilon/2$ and 
$P_S(e)=1-\epsilon$. One wishes to quantize the source with the constraint that neither $0$ nor $1$ may be changed, but $e$ may be quantized to either $0$ or $1$. In other words, the reconstruction is constrained to use only two symbols with the distortion function
\[
d(a,b)=0 \mbox{ if } a=e, \mbox{ or } a=b; \mbox{ and } 1 \mbox{ otherwise.} 
\]
The rate-distortion function for average distortion $D=0$ is given by 
\[
R_{\mbox{BEQ}}(D=0)=\epsilon.
\]
The objective of a covering code is to use as few codewords as possible to cover the space $\{0,1,e\}^n$ with the constraint given above. 
This problem can be considered as the dual of the erasure channel coding problem. In \cite{martinian2003iterative}, a computationally efficient encoding algorithm called iterative quantization was constructed. It may be noted that
this is related but significantly different from the iterative decoding (also called belief propagation, or peeling decoding) used for channel codes. 

A code $\mathbb{C}_2$ of length $n$ is said to cover erasure pattern $E \subset \{1,2,\ldots,n\}$, if for 
any vector $r \in \mathbb{F}_2^{|E|}$, there is a codeword $c_2$ with $c_{2,E}=r$, i.e., the codeword matches $r$ in the said positions in $E$.

\subsection{Quantum Nested Codes and Encoding}

In CSS codes,  both the constituent codes $\mathbb{C}_1$ and $\mathbb{C}_2^{\perp}$ are good channel codes, and decoding happens on these codes $\mathbb{C}_1$ and $\mathbb{C}_2^{\perp}$. 
Although this approach (i.e., designing a good packing code $\mathbb{C}_2^\perp$ instead of a good covering code $\mathbb{C}_2$) leads to capacity-achieving codes for the quantum erasure channel, 
due to a well-established duality~\cite{martinian2003iterative},
this is not true for general channels.
In this paper we adopt the viewpoint that $\mathbb{C}_1$ is a good channel code (packing code), and $\mathbb{C}_2$ is a good source code (covering code).
The decoding happens on these codes.
The motivation comes from the study of quantum capacity of quantum channels~\cite{lloyd1997capacity,shor2002quantum,Devetak2005} as elucidated in the previous subsection.

Consider two nested binary linear codes
\begin{equation}
    \cC_2 \subseteq \cC_1\subseteq\cF_2^n,
\end{equation}
with dimensions $k_2$ and $k_1$, respectively, where
$k_2\leq k_1$. Denote their rates by
$R_i=k_i/n$, for $i\in\{1,2\}$.
The corresponding quantum code encodes $k=(k_1-k_2)$ logical qubits into $n$ code-variable qubits, with quantum code rate
$R_Q=\frac{k}{n}=R_1-R_2$.
Let $G_2\in\cF_2^{k_2\times n}$ be a full-row-rank generator matrix of $\cC_2$, and let $R\in\cF_2^{k\times n}$ contain linearly independent coset representatives of $\cC_1/\cC_2$. Thus,
\begin{equation}
    \cC_1=\operatorname{rowspan}
    \begin{pmatrix}
        R\\G_2
    \end{pmatrix}.
\end{equation}
Each code $\cC_i$, $i\in\{1,2\}$, admits a linear constraint representation 
$\cC_i=\{c\in\cF_2^n:\exists\,s\in\cF_2^{\ell_i},\ s A_i+cB_i=0\}$, 
where $A_i\in\cF_2^{\ell_i\times m_i}$ is a matrix of full row rank and $B_i\in\cF_2^{n\times m_i}$. Here, $c_i$'s denote the 
code variables, and $s_i$'s denote state variables that are not transmitted. The corresponding bipartite graph has code-variable nodes,  state-variable nodes, and check nodes for the scalar constraints, with edges specified by nonzero coefficients of these matrices. 
A parity-check representation is the special case with no state variables and $B_i=H_i^T$. A generator matrix representation is another special case with $A_i=G_i$ and $B_i=I_n$.
Such representations accommodate classical sparse graphical constructions with internal state variables~\cite{MacKay1999,Pfister2004IRA,HsAn10}, as well as recent nested CSS code constructions~\cite{Kasai2026,mercanoglu2026achievabilitycsscodesperformance,kasai2026spatiallycoupledmackaynealhsuanastasopouloscss}.

For a logical basis vector $\ket{u}$,  $u\in\cF_2^k$, the corresponding 
encoded state
is
\begin{equation}
\label{eq:css-encoding}
    \ket{\bar u}
    =
    \frac{1}{\sqrt{2^{k_2}}}
    \sum_{c_2\in \cC_2}
    \ket{uR+c_2}.
\end{equation}

For an arbitrary logical state
\begin{equation}
    \ket{\psi_L}
    =
    \sum_{u\in\cF_2^k}
    \alpha_u\ket{u},
    \qquad
    \sum_u|\alpha_u|^2=1,
\end{equation}
the encoded state is
\begin{equation}
\label{eq:encoded-state}
    \ket{\psi}
    =
    \frac{1}{\sqrt{2^{k_2}}}
    \sum_{u\in\cF_2^k}
    \sum_{c_2\in \cC_2}
    \alpha_u\ket{uR+c_2},
\end{equation}
where all operations in the expression $uR+c_2$ are binary operations performed over $\cF_2$.

\begin{example}[Steane {$[[7,1,3]]$} code]

Throughout this paper, we use the Steane $[[7,1,3]]$ CSS code to illustrate the proposed decoder. We take $\cC_2$ to be the binary $[7,3,4]$ simplex code and $\cC_1=\cC_2^\perp$ to be the $[7,4,3]$ Hamming code. Thus, $k_1=4$, $k_2=3$, and $k=1$.
A  systematic generator matrix for $\cC_2$, together with a coset representative, is
\begin{equation}
\label{eq:steane-matrices}
G_2=
\begin{bmatrix}
1&1&1&0&1&0&0\\
1&1&0&1&0&1&0\\
1&0&1&1&0&0&1
\end{bmatrix},
\qquad
R=
\begin{bmatrix}
1&1&1&1&1&1&1
\end{bmatrix}.
\end{equation}
For this example, $G_2$ also serves as a parity-check matrix for $\cC_1$. 
Note that $\mathbb{C}_1$ is able to correct all erasure patterns  $E$ with $|E|\leq 2$, and $\mathbb{C}_2$ can cover all erasure patterns $E$ with $|E|\leq 2$.

Consider an arbitrary logical qubit
\begin{equation}
    \ket{\psi_L}
    =\alpha\ket0+\beta\ket1,
    \qquad |\alpha|^2+|\beta|^2=1.
\end{equation}

Its encoded state is
\begin{subequations}
\begin{align}
\label{eq:steane-encoded}
\ket{\psi}
&=
\frac{1}{\sqrt8}
\sum_{x\in\mathbb F_2^3}
\left(
\alpha\ket{xG_2}
+\beta\ket{R+xG_2}
\right)\\
&=
\frac{\alpha}{\sqrt8}\Bigl(
\ket{0000000}
+\ket{1011001}
+\ket{1101010}
+\ket{0110011}
\notag\\
&\qquad\qquad
+\ket{1110100}
+\ket{0101101}
+\ket{0011110}
+\ket{1000111}
\Bigr)
\notag\\
&\quad+
\frac{\beta}{\sqrt8}\Bigl(
\ket{1111111}
+\ket{0100110}
+\ket{0010101}
+\ket{1001100}
\notag\\
&\qquad\qquad
+\ket{0001011}
+\ket{1010010}
+\ket{1100001}
+\ket{0111000}
\Bigr).
\label{eq:steane-encoded-expanded}
\end{align}
\end{subequations}

\end{example}

\subsection{Encoded State After Erasure}

With this background, we construct a coding scheme in which the denser code $\mathbb{C}_1$ is a good channel code for the classical binary erasure channel with finite graphical complexity, so it is amenable to iterative decoding such as belief propagation. The lighter code 
$\mathbb{C}_2$ is a good covering code for the binary erasure source quantization problem with finite graphical complexity, so it is amenable to iterative quantization algorithms.

Let $E\subseteq\{1,\ldots,n\}$ denote the set of erased physical positions and let $\bar E$ denote its complement. For a binary vector $c$, write $c_E$ and $c_{\bar E}$ for its restrictions to the corresponding coordinate sets.

Conditioned on the observed erasure pattern $E$, the received state is
\begin{equation}
    \rho_E(\psi)=
    \left(\ket e\bra e\right)^{\otimes |E|}
    \otimes
    \operatorname{Tr}_E
    \left(\ket\psi\bra\psi\right).
\end{equation}

For a fixed erasure pattern $E$, the code-variable qubits indexed by $E$ are replaced by the erasure symbol $\ket e$. The surviving density operator is obtained by tracing out the erased subsystem.
One can perform the partial tracing explicitly to arrive at the following expression for $\rho_E(\psi)$
\begin{subequations}
\label{eq:erasure-branches}
\begin{align}
    \rho_E(\psi)
    &=     \sum_{b\in\mathbb F_2^{|E|}}
    \ket{\psi_b}\bra{\psi_b},\\
    \ket{\psi_b}
    &=
    \ket{e_E}\otimes
    \frac{1}{\sqrt{2^{k_2}}}
    \sum_{u\in\mathbb F_2^k}
    \alpha_u
    \sum_{\substack{c_2\in\cC_2:\\
    uR_E+c_{2,E}=b}}
    \ket{uR_{\bar E}+c_{2,\bar E}}.
\end{align}
\end{subequations}

where the tensor factors are ordered with the erased subsystem first for notational convenience. The cross terms between different erased values $b\neq b'$ disappear because the erased subsystem has been traced out. Consequently, the receiver has a mixture of branches rather than their coherent superposition.
Throughout the paper, branch vectors are left unnormalized. For fixed $E$, the vector $\ket{\psi_b}$ has squared norm $p_{b|E}:=\langle\psi_b|\psi_b\rangle$, with $\sum_b p_{b|E}=1$. Thus, the branch weights are already included in $\rho_E(\psi)=\sum_b\ket{\psi_b}\bra{\psi_b}$. The subsequent recovery operations preserve these weights; we do not renormalize individual branches.
Importantly, the receiver does not have access to the branch label $ b$. The only classical information obtained from the erasure channel is the erasure pattern $E$.
Nevertheless, the logical quantum information may remain recoverable. The objective of the decoder is not to restore coherence between different erased-subsystem branches, but to extract the original logical state coherently from either nonzero branch, leaving a residual state independent of the logical input.

\begin{example*}[Steane code: erasure of qubits 1 and 2]

Consider the erasure pattern $E=\{1,2\}$. The two erased positions are replaced by $\ket e$, while qubits $3$ through $7$ remain unchanged. The receiver learns $E$, but not the original binary values $b=(b_1,b_2)$ of the erased qubits.

For the matrix in~\eqref{eq:steane-matrices},
\begin{equation}
\label{eq:steane-restricted}
R_E=\begin{bmatrix}1&1\end{bmatrix}.
\end{equation}

There are four possible inaccessible branches, corresponding to $b\in\mathbb F_2^2$. The unnormalized branch states in~\eqref{eq:erasure-branches} are
\begin{subequations}
\label{eq:steane-erased-branches}
\begin{align}
\label{eq:steane-erased-branch}
\ket{\psi_{00}}
=&\frac{\ket{ee}}{\sqrt8}\otimes\Big[
\alpha(\ket{00000}+\ket{11110})
+\beta(\ket{10101}+\ket{01011})
\Big]\\
\ket{\psi_{01}}
=&\frac{\ket{ee}}{\sqrt8}\otimes\Big[
\alpha(\ket{10011}+\ket{01101})
+\beta(\ket{00110}+\ket{11000})
\Big]\\
\ket{\psi_{10}}
=&\frac{\ket{ee}}{\sqrt8}\otimes\Big[
\alpha(\ket{11001}+\ket{00111})
+\beta(\ket{01100}+\ket{10010})
\Big]\\
\ket{\psi_{11}}
=&\frac{\ket{ee}}{\sqrt8}\otimes\Big[
\alpha(\ket{01010}+\ket{10100})
+\beta(\ket{11111}+\ket{00001})
\Big].
\end{align}
\end{subequations}


The receiver does not observe $b$, so the decoding process must work for every branch using only the known erasure pattern $E$.

\end{example*}

\section{Architecture I: Classical Decoding/Quantization with Coherent Quantum Reconstruction and Cleanup}
\label{sec:classical-architecture}

\begin{figure}[htbp]
    \centering
    \includegraphics[width=\columnwidth]{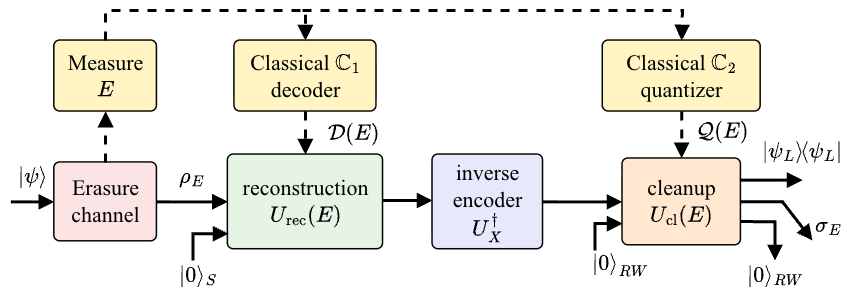}
    \caption{Classically scheduled coherent erasure recovery: variable reconstruction, inverse binary encoding, and coherent quantization-based cleanup.}
    \label{fig:architecture1}
\end{figure}

We consider the classically scheduled coherent erasure decoder shown in Fig.~\ref{fig:architecture1}. The receiver measures the erasure flags to determine $E$, without measuring the computational-basis values of the surviving qubits. Classical processing prepares a reconstruction schedule $\mathcal{D}(E)$ for $\cC_1$ and a quantization reservation schedule $\mathcal{Q}(E)$ for $\cC_2$. The receiver then performs the following three quantum operations, which are described in the sequel: coherent symbol reconstruction, inverse binary encoding, and coherent quantization-based cleanup. 

\subsection{Classical Erasure Decoding for \texorpdfstring{$\cC_1$}{C1}}
\label{subsec:classical-decoding}
After erasure, the receiver retains the surviving coordinates of each computational-basis codeword. The receiver initializes the erased registers and one working qubit per state variable to $\ket0$.  We seek a coherent reconstruction operation satisfying

\begin{equation}
\label{eq:required-reconstruction-action}
    U_{\mathrm{rec}}(E):
    \quad
    \ket{0}_E\ket{c_{\bar E}}_{\bar{E}}\ket{0}_S
    \longmapsto
    \ket{c_E}_E\ket{c_{\bar E}}_{\bar{E}}\ket{s}_S,
    \qquad sA_1+cB_1=0.
\end{equation}

The same operation must act on every codeword without measuring the surviving binary values. It therefore reconstructs the codewords coherently within each erasure-environment branch. Such an operation requires the surviving coordinates to determine the erased coordinates uniquely. For a received vector $y$ on $\bar E$, erasure decoding seeks a unique vector $(s,c_1)$ satisfying

\begin{equation}
\label{eq:codeword-decoding-condition}
    sA_1+c_1 B_1=0,
    \qquad c_{1,\bar E}=y.
\end{equation}
Partitioning $B_1$ according to the erased and non-erased rows the above is equivalent to solving the linear system of equations
\begin{equation}
\label{eq:classical-decoding-system}
    [s \ c_{1,E}] \begin{bmatrix} A_1 \\ B_{1,E} \end{bmatrix}
    = y B_{1,\bar{E}}.
\end{equation}

The classical decoder determines a schedule of binary operations for recovering the erased coordinates, or declares failure. For example, Gaussian elimination in~\eqref{eq:classical-decoding-system} provides such a schedule. 
In the following, we describe a suboptimal decoder, the peeling decoder that attempts to solve~\eqref{eq:classical-decoding-system}.
The operations are prepared classically without evaluating the unknown quantum-register values and are executed coherently as described in Section~\ref{subsec:coherent-reconstruction}.

\paragraph{Peeling as a suboptimal decoding procedure.}
For a graph $\mathcal{G}_1$, 
characterizing the sparse structure of $\cC_1$, the peeling decoder~\cite[Sec.~3.19]{Richardson_Urbanke_2008}, given in Algorithm~\ref{alg:classical-peeling}, constructs an ordered reconstruction schedule
\begin{equation}
\label{eq:classical-peeling-schedule}
    \mathcal{D}(E)=\big((h_1,v_1),\ldots,(h_t,v_t)\big).
\end{equation}

\begin{algorithm}[htbp]
\caption{Classical Peeling Schedule for $\cC_1$}
\label{alg:classical-peeling}
\begin{algorithmic}[1]
\Require Graph $\mathcal{G}_1$, erasure set $E$
\Ensure $(\mathrm{SUCCESS},\mathcal{D}(E))$
        or $\mathrm{FAILURE}$

\State $F\gets E\cup S$, $\mathcal{D}(E)\gets[\,]$, $\mathcal{L}\gets\varnothing$
 
\For{each check $h$}
    \Comment{List of checks with remaining degree 1}
    \State $d_h^{\mathrm{rem}}\gets|N(h)\cap\mathcal{F}|$
        \If{$d_h^{\mathrm{rem}}=1$}
            \State $\mathcal{L}\gets\mathcal{L}\cup \{h\}$
        \EndIf
\EndFor

\While{$F\neq\varnothing$}
    \If{$\mathcal{L}=\varnothing$}
        \State \Return $\mathrm{FAILURE}$
    \EndIf

    \State Pop check $h$ from $\mathcal{L}$
    \If{$d_h^{rem}=1$}
        \State Let $\{v\}=N(h)\cap F$
        \State Append $(h,v)$ to $\mathcal{D}(E)$
        \State $F\gets F\setminus\{v\}$

        \For{each check $h'\in N(v)$}  \Comment{$N(v)$ is the set of check nodes connected to $v$}
            \State $d_{h'}^{rem}\gets d_{h'}^{rem}-1$
            \If{$d_{h'}^{rem}=1$}
                \State Push $h'$ into $\mathcal{L}$
            \EndIf
        \EndFor
    \EndIf
\EndWhile

\State \Return $(\mathrm{SUCCESS},\mathcal{D}(E))$
\end{algorithmic}
\end{algorithm}

Each pair $(h_\ell,v_\ell)$ identifies a check $h_{\ell}$ whose unique unresolved neighbor is $v_\ell$. The schedule depends only on the graph and the erasure pattern; the corresponding XOR assignments are evaluated coherently in Section~\ref{subsec:coherent-reconstruction}. Peeling succeeds when every erased code variable and every state variable has been scheduled and declares failure if no eligible check remains while erasures are unresolved. The scheduling complexity is established in Lemma~\ref{lem:reconstruction-complexity}.

We note that Algorithm~\ref{alg:classical-peeling} applies to graphs whose variable nodes represent both the code variables, $c_1$, as well as the state variables, $s$ (if they exist). In the latter case, the erasure pattern is understood as the union of the channel erasure pattern with the state nodes. 

\begin{example*}[Steane example: classical peeling schedule.]
Continuing the Steane example with $E=\{1,2\}$, we construct the reconstruction schedule using the Tanner graph defined by the parity-check matrix in~\eqref{eq:steane-matrices}. Algorithm~\ref{alg:classical-peeling} initially selects check $3$, whose unique unresolved neighbor is variable $1$, and records $(3,1)$. After removing variable $1$ from the unresolved set, check $1$ has exactly one unresolved neighbor, variable $2$, giving the next pair $(1,2)$.

The reconstruction schedule is therefore
\begin{equation}
\label{eq:steane-peeling}
    \mathcal{D}(\{1,2\})
    =
    \big((3,1),(1,2)\big).
\end{equation}
Both erased variables have been scheduled, so peeling succeeds. Figure~\ref{fig:steane_peeling} shows the two scheduling steps. No binary values are evaluated at this stage; Section~\ref{subsec:coherent-reconstruction} implements the scheduled reconstruction operations coherently.

\begin{figure}[htbp]
    \centering
    \includegraphics[width=0.8\columnwidth]{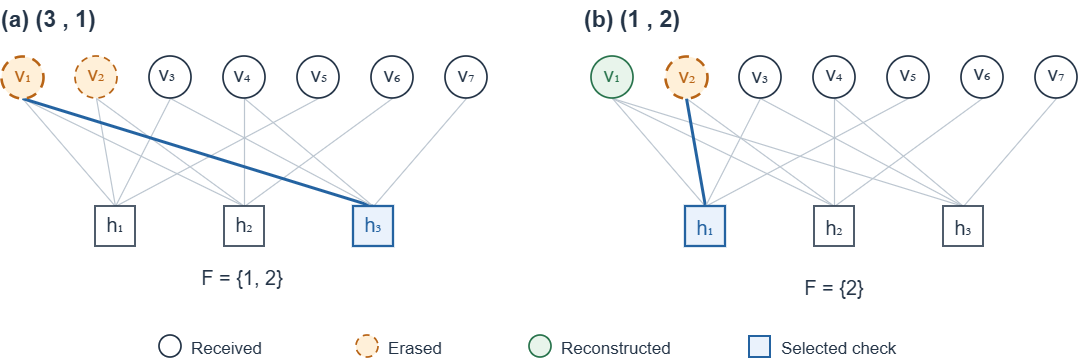}
    \caption{Classical peeling schedule for the running Steane example. Panel (a) selects $(3,1)$; panel (b) selects $(1,2)$ after removing variable $1$ from the unresolved set. Each pair lists the check index followed by the variable index. Orange dashed circles denote unresolved erased variables, green circles denote variables already scheduled for recovery, and blue highlighting identifies the selected check--variable pair. No binary values are evaluated.}
    \label{fig:steane_peeling}
\end{figure}

\end{example*}

\subsection{Coherent Physical-Symbol Reconstruction via Classically Scheduled Gates}
\label{subsec:coherent-reconstruction}

The receiver implements the operations prepared in Section~\ref{subsec:classical-decoding} using CNOT gates controlled by surviving or previously reconstructed symbols. 

Consider the registers $Y$ holding the code variables together with auxiliary registers, $S$,  holding the state variables (if present). 
For the following scheduled implementation, the receiver uses $a=0$ and $a'=0$ in~\eqref{eq:coherent-reconstruction-action}. 
The resulting reconstruction operation satisfies

\begin{equation}
\label{eq:coherent-reconstruction-action}
\begin{aligned}
    U_{\mathrm{rec}}(E):\quad
    &\ket{a}_E\ket{c_{\bar E}}_{\bar E}\ket{a'}_S\\
    &\quad\longmapsto
    \ket{a+c_E}_E\ket{c_{\bar E}}_{\bar E}\ket{a'+s}_S,
    \qquad s A_1+c B_1=0.
\end{aligned}
\end{equation}

We now implement $U_{\mathrm{rec}}(E)$ using the classical peeling schedule in \eqref{eq:classical-peeling-schedule}. For each pair $(h_\ell,v_\ell)$ in the schedule~\eqref{eq:classical-peeling-schedule}, the quantum gate
\begin{equation}
\begin{aligned}
    &\ket{x_{N(h_\ell)\setminus\{v_\ell\}}}
    \ket0_{v_\ell}\\
    &\quad\longmapsto
    \ket{x_{N(h_\ell)\setminus\{v_\ell\}}}
    \ket{
    \bigoplus_{i\in N(h_\ell)\setminus\{v_\ell\}}x_i
    }_{v_\ell}.
\end{aligned}
\end{equation}
is applied, where each $x_i$ refers to either a code-variable register or a state-variable register. The peeling schedule guarantees that every register which serves as a control in~\eqref{eq:classical-peeling-schedule}  is either unerased or has already been reconstructed. When its reconstruction operation is applied, the target $v_\ell$ is still in $\ket0$. Each recovered variable is subsequently used only as a control. Algorithm~\ref{alg:coherent-reconstruction} summarizes the previous reconstruction steps. It recovers the erased code variables and state variables.
The update $y_j \leftarrow y_j \oplus y_i$ is implemented by the CNOT gate $\mathrm{CNOT}_{i\to j}$.

\begin{algorithm}[t]
\caption{Coherent Reconstruction Using a Peeling Schedule}
\label{alg:coherent-reconstruction}
\begin{algorithmic}[1]
\Require Successful schedule $\mathcal D(E)=((h_1,v_1),\ldots,(h_t,v_t))$; code-variable registers $Y=(Y_E,Y_{\bar E})$, with surviving code variables in $Y_{\bar E}$; replacement registers $Y_E$ and state-variable registers $S$ initialized to $\ket0$
\Ensure Reconstructed code variables $c$ in $Y$ and state variables $s$ in $S$, satisfying $sA_1+cB_1=0$
\For{$\ell=1,\ldots,t$}
    \For{$i\in N(h_\ell)\setminus\{v_\ell\}$}
        \State Apply $\operatorname{CNOT}_{i\to v_\ell}$
    \EndFor
\EndFor
\end{algorithmic}
\end{algorithm}

\begin{lemma}[Reconstruction complexity]
\label{lem:reconstruction-complexity}
Let $\mathcal G_1$ be the constraint graph defined by $sA_1+cB_1=0$, with $n_v=n+\ell_1$ variable nodes, $n_h=m_1$ check nodes, and $M=\operatorname{nnz}(A_1)+\operatorname{nnz}(B_1)$ edges. Here, $\operatorname{nnz}(\cdot)$ denotes the number of nonzero matrix entries.  


Algorithm~\ref{alg:classical-peeling} takes $O(n_v+n_h+M)$ time, whether it returns success or failure. If it succeeds, producing $\mathcal D(E)=((h_1,v_1),\ldots,(h_t,v_t))$, Algorithm~\ref{alg:coherent-reconstruction} uses $O(M)$ CNOT gates.
Furthermore, if $A_1$ and $B_1$ have $O(n)$ nonzero entries in total and there are $O(n)$ state variables and checks, these bounds reduce to $O(n)$ classical time and $O(n)$ CNOT gates, respectively.
\end{lemma}

\begin{proof}
First consider the time complexity of Algorithm~\ref{alg:classical-peeling}. Use adjacency lists, constant-time membership flags for $F$, and constant-time queue operations. Initializing the unresolved set $F$, the empty schedule, and the empty queue in line~1 takes $O(n_v)$ time. Computing the unresolved-neighbor counts and initializing the queue in lines~2--7 takes $O(n_h+M)$ time. Each variable is resolved at most once, so the neighboring-check updates in lines~17--22 traverse at most $M$ edges in total. Finding the unique unresolved neighbor in line~14 requires scanning the selected check's adjacency list; each check is selected at most once, so these scans also take $O(M)$ time in total. Each check enters the queue at most once because its unresolved-neighbor count only decreases, giving $O(n_h)$ queue operations. These observations give the stated $O(n_v+n_h+M)$ scheduling bound.

Now consider the gate complexity of Algorithm~\ref{alg:coherent-reconstruction}. Each selected check contributes $|N(h_\ell)|-1$ CNOT gates and is selected at most once. The total gate count is therefore at most $M$.

\end{proof}

By linearity, the same circuit reconstructs the codewords coherently within each erasure-environment branch. For each $c\in\cC_1$, let $s(c)$ denote the unique state vector satisfying $s(c)A_1+cB_1=0$, where uniqueness follows from the full row rank of $A_1$.
Since the branch label $b$ is inaccessible, the joint reconstructed state of $Y$ and $S$ is

\begin{subequations}
\begin{align}
\label{eq:reconstructed-physical-density}
    \rho_{\mathrm{rec},E}(\psi)
    &=
    \sum_{b\in\cF_2^{|E|}}
    \ket{\psi'_b}\bra{\psi'_b},\\
\label{eq:reconstructed-physical-branch}
    \ket{\psi'_b}
    &=
    \frac{1}{\sqrt{2^{k_2}}}
    \sum_{u\in\cF_2^k}
    \alpha_u
    \sum_{\substack{c_2\in\cC_2:\\
    uR_E+c_{2,E}=b}}
    \ket{uR+c_2}\ket{s(uR+c_2)}_S.
\end{align}
\end{subequations}

Thus, successful reconstruction restores the computational-basis codewords within each branch, but does not generally restore the original encoded pure state. The jointly reconstructed density operator remains a mixture over the inaccessible erasure-environment branches.

\begin{example*}[Steane example: coherent reconstruction.]
For the schedule in~\eqref{eq:steane-peeling}, the first operation reconstructs qubit $1$ using check $3$, whose support is $\{1,3,4,7\}$. The corresponding operation is
\begin{equation}
\label{eq:steane-patch1}
    U_{3,1}
    =
    \operatorname{CNOT}_{3\to1}
    \operatorname{CNOT}_{4\to1}
    \operatorname{CNOT}_{7\to1}.
\end{equation}
The second operation reconstructs qubit $2$ using check $1$, whose support is $\{1,2,3,5\}$:
\begin{equation}
\label{eq:steane-patch2}
    U_{1,2}
    =
    \operatorname{CNOT}_{1\to2}
    \operatorname{CNOT}_{3\to2}
    \operatorname{CNOT}_{5\to2}.
\end{equation}
Thus, the reconstruction circuit is
\begin{equation}
    U_{\mathrm{rec}}(E)
    =
    U_{1,2}U_{3,1}.
\end{equation}
After initializing the erased qubits to $\ket0$, these six CNOT gates coherently reconstruct the two missing code variables. Figure~\ref{fig:coherent_reconstruction} shows the two groups of reconstruction gates. In Fig.~\ref{fig:coherent_reconstruction}, $Y_i$ denotes code-variable qubit register $i$.

\begin{figure}[htbp]
    \centering
    \includegraphics[height=5cm]{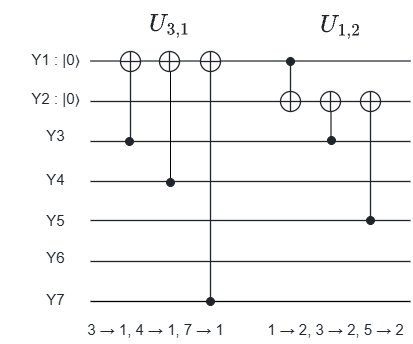}
    \caption{Coherent reconstruction corresponding to Fig.~\ref{fig:steane_peeling}. Panel (a) reconstructs qubit~1 using controls $3,4,7$; panel (b) subsequently reconstructs qubit~2 using controls $1,3,5$. Each erased target is initialized in $|0\rangle$ before reconstruction.}
    \label{fig:coherent_reconstruction}
\end{figure}

Applying the reconstruction circuit to the four erasure branches in~\eqref{eq:steane-erased-branch} gives the following unnormalized states
\begin{subequations}
\begin{align}
\label{eq:steane-reconstructed}
\ket{\psi'_{00}}&=\frac{1}{\sqrt8}\Big[
    \alpha(\ket{0000000}+\ket{0011110}) 
    +\beta(\ket{0010101}+\ket{0001011})\Big] \\
\ket{\psi'_{01}}&=\frac{1}{\sqrt8}\Big[
    \alpha(\ket{0110011}+\ket{0101101}) 
    +\beta(\ket{0100110}+\ket{0111000})\Big] \\
\ket{\psi'_{10}}&=\frac{1}{\sqrt8}\Big[
    \alpha(\ket{1011001}+\ket{1000111}) 
    +\beta(\ket{1010010}+\ket{1001100})\Big] \\
\ket{\psi'_{11}}&=\frac{1}{\sqrt8}\Big[
    \alpha(\ket{1101010}+\ket{1110100}) 
    +\beta(\ket{1111111}+\ket{1100001})\Big].
\end{align}
\end{subequations}
In each branch, every reconstructed code string has restriction $b$ on the erased coordinates $E=\{1,2\}$. Each branch has squared norm $1/4$. The same circuit acts on every inaccessible branch without requiring knowledge of the erased binary values.

\end{example*}

\subsection{Inverse Binary Encoding}
\label{subsec:inverse-encoding}

Successful reconstruction restores the variables within each inaccessible branch. We first apply the inverse binary encoder to expose the logical register, and then remove its residual dependence on the auxiliary register by a classically prepared cleanup operation. Since $G_2$ has full row rank, every $c_2\in\cC_2$ has a unique representation $c_2=xG_2$, with $x\in\cF_2^{k_2}$. The reconstructed branch can therefore be written as

\begin{equation}
\label{eq:reconstructed-generator-branch}
    \ket{\psi'_b}
    =\frac{1}{\sqrt{2^{k_2}}}
    \sum_{u\in\cF_2^k}\alpha_u
    \sum_{\substack{x\in\cF_2^{k_2}:\\uR_E+xG_{2,E}=b}}
    \ket{uR+xG_2}\ket{s(uR+xG_2)}_S.
\end{equation}

Here $G_{2,E}$ contains the columns of $G_2$ indexed by $E$. 
\
Consider the full encoder transformation whose action on initialized registers is
\begin{equation}
\label{eq:full-encoder-action}
\begin{aligned}
U_X:\quad
    &\ket{u}_L\ket{x}_X
    \ket{0}_Z^{\otimes(n-k_1)}
    \ket{0}_S^{\otimes\ell_1}\\
    &\quad\longmapsto
    \ket{uR+xG_2}_Y\ket{s(uR+xG_2)}_S,
\end{aligned}
\end{equation}
where registers $L$, $X$, $Z$, and $S$ contain $(k_1-k_2)$, $k_2$, $(n-k_1)$, and $\ell_1$ qubits, respectively.

The inverse encoder satisfies

\begin{equation}
\label{eq:inverse-encoder-action}
    U_{\mathrm X}^{\dagger}\ket{uR+xG_2}\ket{s(uR+xG_2)}_S
    =\ket u_L\ket x_X\ket0^{\otimes(n+\ell_1-k_1)}.
\end{equation}

The rows of $R$ together with those of $G_2$, augmented by their corresponding state vectors, form a (dense) basis of the 
code $\{(c,s):sA_1+cB_1=0\}$. Extending this basis to $\cF_2^{n+\ell_1}$ defines an invertible binary transformation, implementable using CNOT gates on the existing $n+\ell_1$ code-variable qubits and state-variable qubits. The $k_2$ auxiliary qubits and the $(n+\ell_1-k_1)$ zero qubits are part of these registers. An arbitrary invertible binary transformation on $n+\ell_1$ qubits can be implemented using $O((n+\ell_1)^2)$ CNOT gates, giving an $O(n^2)$ worst-case bound for inverse encoding when $\ell_1=O(n)$.

Applying inverse encoding to each branch
in~\eqref{eq:reconstructed-generator-branch} gives the
unnormalized branch
\begin{equation}
\label{eq:post-inverse-branch}
    \ket{\chi_b}
    =\frac{1}{\sqrt{2^{k_2}}}
    \sum_u\alpha_u
    \sum_{\substack{x\in\cF_2^{k_2}:\\uR_E+xG_{2,E}=b}}
    \ket{u}_L\ket{x}_X
    \ket{0}_Z^{\otimes(n-k_1)}
    \ket{0}_S^{\otimes\ell_1}.
\end{equation}
Although $u$ now occupies the logical register, the allowed auxiliary labels still depend on $u$ through the branch constraint $uR_E+xG_{2,E}=b$. Inverse encoding alone therefore need not recover the original logical state. Sections~\ref{subsec:c2-quantization} and~\ref{subsec:coherent-cleanup} prepare and implement the cleanup operation that removes this dependence without measuring either $u$ or $b$.

\begin{example*}[Steane example: inverse encoding.]

For the systematic generator $G_2=[P\mid I_3]$, the coordinates of $c_2=xG_2$ are $x=(c_{2,5},c_{2,6},c_{2,7})$. Applying inverse encoding to the four reconstructed branches gives the following states, where the final $\ket{000}$ factor is omitted:
\begin{subequations}
\begin{align}
\label{eq:steane-inverse-branch}
\ket{\chi_{00}}
&=\frac{1}{\sqrt8}\Big[
\alpha\ket0_L(\ket{000}+\ket{110})_X
+\beta\ket1_L(\ket{100}+\ket{010})_X\Big],\\
\label{eq:steane-inverse-branch-01}
\ket{\chi_{01}}
&=\frac{1}{\sqrt8}\Big[
\alpha\ket0_L(\ket{011}+\ket{101})_X
+\beta\ket1_L(\ket{001}+\ket{111})_X\Big],\\
\label{eq:steane-inverse-branch-10}
\ket{\chi_{10}}
&=\frac{1}{\sqrt8}\Big[
\alpha\ket0_L(\ket{001}+\ket{111})_X
+\beta\ket1_L(\ket{011}+\ket{101})_X\Big],\\
\label{eq:steane-inverse-branch-11}
\ket{\chi_{11}}
&=\frac{1}{\sqrt8}\Big[
\alpha\ket0_L(\ket{010}+\ket{100})_X
+\beta\ket1_L(\ket{000}+\ket{110})_X\Big].
\end{align}
\end{subequations}
Note how, in every branch, the auxiliary superpositions accompanying the two logical values (X ancillas) differ, so the logical state cannot be extracted coherently at this stage. Cleanup must make them identical using only the known erasure pattern and the fixed code and encoder.

\end{example*}

\subsection{Classical Erasure Quantization}
\label{subsec:c2-quantization}

After inverse encoding, the logical and auxiliary registers remain coupled through the constraint $uR_E+xG_{2,E}=b$ in~\eqref{eq:post-inverse-branch}. We seek a cleanup transformation
\begin{equation}
\label{eq:cleanup-required-action}
    \ket u_L\ket x_X
    \longmapsto
    \ket u_L\ket{x+g_E(u)}_X,
\end{equation}
that disentangles the logical register from the auxiliary register. 
Indeed, applying this transformation to~\eqref{eq:post-inverse-branch} results in
\begin{equation}
\label{eq:post-inverse-branch-transformed}
    \ket{\chi'_b}
    =\frac{1}{\sqrt{2^{k_2}}}
    \sum_u\alpha_u
    \sum_{\substack{x\in\cF_2^{k_2}:\\uR_E+xG_{2,E}=b}}
    \ket u_L\ket{x+g_E(u)}_X\ket{0}_Z^{\otimes(n-k_1)}\ket{0}_S^{\otimes\ell_1}.
\end{equation}
Writing $x'=x+g_E(u)$, the branch constraint becomes
$uR_E+(x'+g_E(u))G_{2,E}=b$.
Thus, it is sufficient to construct a map satisfying
\begin{equation}
\label{eq:cleanup-required-condition}
    g_E(u)G_{2,E}=uR_E,
\end{equation}
so that the auxiliary constraint reduces to $x'G_{2,E}=b$, independent of the logical label $u$. With this choice~\eqref{eq:post-inverse-branch-transformed} simplifies to 
\begin{align}
\label{eq:post-inverse-branch-simplified}
    \ket{\chi'_b}
    &=\frac{1}{\sqrt{2^{k_2}}}
    \sum_u\alpha_u
    \sum_{\substack{x'\in\cF_2^{k_2}:\\x'G_{2,E}=b}}
    \ket u_L\ket{x'}_X\ket{0}_Z^{\otimes(n-k_1)}\ket{0}_S^{\otimes\ell_1} \\
    &=\frac{1}{\sqrt{2^{k_2}}}
    \left( \sum_u\alpha_u \ket u_L \right)
    \left( \sum_{\substack{x'\in\cF_2^{k_2}:\\x'G_{2,E}=b}}
    \ket{x'}_X \right)
    \ket{0}_Z^{\otimes(n-k_1)}
    \ket{0}_S^{\otimes\ell_1},
\end{align}
thus separating the logical state from the auxiliary ancillas.

Since $g_E(u)G_2\in\cC_2$, condition~\eqref{eq:cleanup-required-condition} asks for a codeword of $\cC_2$ whose restriction to $E$ equals $uR_E$. This is a binary erasure-quantization problem~\cite{martinian2003iterative}, with the coordinates on $E$ prescribed and those on $\bar E$ unconstrained. 
We remark that $\cC_2$ serves as a good covering code at this stage of the decoding process, i.e., it enables disentangling the logical state from the remaining registers!
We next describe the classical implementation of a graph-based quantizer. Section~\ref{subsec:coherent-cleanup} describes its coherent implementation.

%
Such an operation requires each prescribed target to extend to a codeword. For a prescribed vector \(r\) on \(E\), (distortionless) erasure quantization seeks a vector $(s_2,c_2)$ satisfying

\begin{equation}
\label{eq:codeword-cleanup-condition}
    s_2 A_2 + c_2 B_2=0,
    \qquad c_{2,E}=r.
\end{equation}
Partitioning $B_2$ according to the erased and non-erased rows the above is equivalent to solving the linear system of equations
\begin{equation}
\label{eq:codeword-cleanup-condition1}
    [s_2 \ c_{2,\bar{E}}] \begin{bmatrix} A_2 \\ B_{2,\bar{E}} \end{bmatrix}
    = r B_{2,E}.
\end{equation}

A classical quantizer determines a schedule of binary operations for completing the prescribed coordinates, or declares failure. For example, Gaussian elimination in~\eqref{eq:codeword-cleanup-condition1}, with free variables set to zero, provides such a schedule when the required target admits a solution. 
In the following, we describe a suboptimal quantizer, the peeling quantizer that attempts to solve~\eqref{eq:codeword-cleanup-condition1}.
We prepare the operations classically without evaluating the unknown quantum-register values and execute them coherently in Section~\ref{subsec:coherent-cleanup}.

\paragraph{Peeling as a suboptimal quantization procedure.}

For the constraint graph $\mathcal G_2$, Algorithm~\ref{alg:erasure-quantization} adapts the reservation principle of Martinian and Yedidia~\cite[Sec.~3.4.2, Table~2]{martinian2003iterative} to the representation $s_2A_2+c_2B_2=0$ and constructs an ordered reservation schedule
\begin{equation}
\label{eq:quantizer-reservation-schedule}
    \mathcal{Q}(E)=\big((v_1,h_1),\ldots,(v_q,h_q)\big).
\end{equation}

\begin{algorithm}[htbp]
\caption{Classical Quantization Reservation Schedule for $\mathcal{C}_2$}
\label{alg:erasure-quantization}
\begin{algorithmic}[1]
\Require Graph $\mathcal{G}_2$,
         prescribed codeword positions $E$
\Ensure $(\mathrm{SUCCESS},\mathcal{Q}(E))$
        or $\mathrm{FAILURE}$

\State $\mathcal{U}\gets
    \{s_a:a=1,\ldots,\ell_2\}\cup\{c_{2,i}:i\notin E\}$
    \Comment{Unprescribed variables}
\State $\mathcal{T}\gets$ all checks,
    $\mathcal{Q}(E)\gets[\,]$,
    $\mathcal{L}\gets\varnothing$
\For{each $v\in\mathcal{U}$}
    \Comment{List of unprescribed variables with remaining degree 1}
    \State $d_v^{\mathrm{rem}}\gets|N(v)\cap\mathcal{T}|$
        \If{$d_v^{\mathrm{rem}}=1$}
            \State $\mathcal{L}\gets\mathcal{L}\cup \{v\}$
        \EndIf
\EndFor

\While{$\mathcal{T}\neq\varnothing$}
    \If{$\mathcal{L}=\varnothing$}
        \State \Return $\mathrm{FAILURE}$
        \Comment{A core or residual constraints remain}
    \EndIf

    \State Pop $v$ from $\mathcal{L}$

    \If{$d_v^{\mathrm{rem}}=1$}
        \State Let $\{h\}=N(v)\cap\mathcal{T}$
        \State Append $(v,h)$ to $\mathcal{Q}(E)$
        \Comment{Keep the original neighborhood of $h$}
        \State $\mathcal{U}\gets\mathcal{U}\setminus\{v\}$
        \State $\mathcal{T}\gets\mathcal{T}\setminus\{h\}$

        \For{each $v'\in N(h)\cap\mathcal{U}$}
            \State $d_{v'}^{\mathrm{rem}}
                \gets d_{v'}^{\mathrm{rem}}-1$
            \If{$d_{v'}^{\mathrm{rem}}=1$}
                \State Push $v'$ into $\mathcal{L}$
            \EndIf
        \EndFor
    \EndIf
\EndWhile

\State \Return $(\mathrm{SUCCESS},\mathcal{Q}(E))$
\end{algorithmic}
\end{algorithm}

Each pair $(v_\ell,h_\ell)$ identifies an unprescribed variable whose unique remaining neighboring constraint is $h_\ell$. The schedule depends only on the graph and the prescribed positions; the corresponding XOR assignments are evaluated coherently in Section~\ref{subsec:coherent-cleanup}. Reservation succeeds when every constraint in the initial set $\mathcal T$ has been reserved and declares failure if no eligible variable remains while constraints in $\mathcal T$ are unresolved. The scheduling complexity is established in Lemma~\ref{lem:cleanup-complexity}.

\begin{example*}[Steane example: quantization reservation schedule]
Continuing the same example, we construct the quantization reservation schedule with coordinates $1$ and $2$ prescribed. We use the graph defined by the local constraints
\begin{subequations}
\begin{alignat}{2}
\label{eq:steane-quantizer-constraints}
    h_1:&\quad c_{2,1}+c_{2,5}+c_{2,6}+c_{2,7} &=&0,\\
    h_2:&\quad c_{2,2}+c_{2,5}+c_{2,6} &=&0,\\
    h_3:&\quad c_{2,3}+c_{2,5}+c_{2,7} &=&0,\\
    h_4:&\quad c_{2,4}+c_{2,6}+c_{2,7} &=&0.
\end{alignat}
\end{subequations}
Figure~\ref{fig:steane_quantization} shows the four scheduling steps. 
Algorithm~\ref{alg:erasure-quantization} initially selects variable $3$, whose unique remaining neighboring constraint is $3$, and records $(3,3)$. After removing constraint $3$, it selects variable $4$ and records $(4,4)$, removing constraint $4$. Variable $7$ then has only constraint $1$ remaining, giving the next pair $(7,1)$. After removing constraint $1$, variable $5$ has only constraint $2$ remaining, giving the final pair $(5,2)$.

The resulting reservation schedule is therefore
\begin{equation}
\label{eq:steane-quantizer-schedule}
    \mathcal{Q}(\{1,2\})
    =
    \big((3,3),(4,4),(7,1),(5,2)\big).
\end{equation}
All constraints have been removed, so reservation succeeds. No constraint involves only prescribed coordinates, and variable $6$ remains unreserved. No binary values are evaluated at this stage; Section~\ref{subsec:coherent-cleanup} implements the scheduled assignments coherently in reverse order.

\begin{figure}[htbp]
    \centering
    \includegraphics[width=0.8\columnwidth]{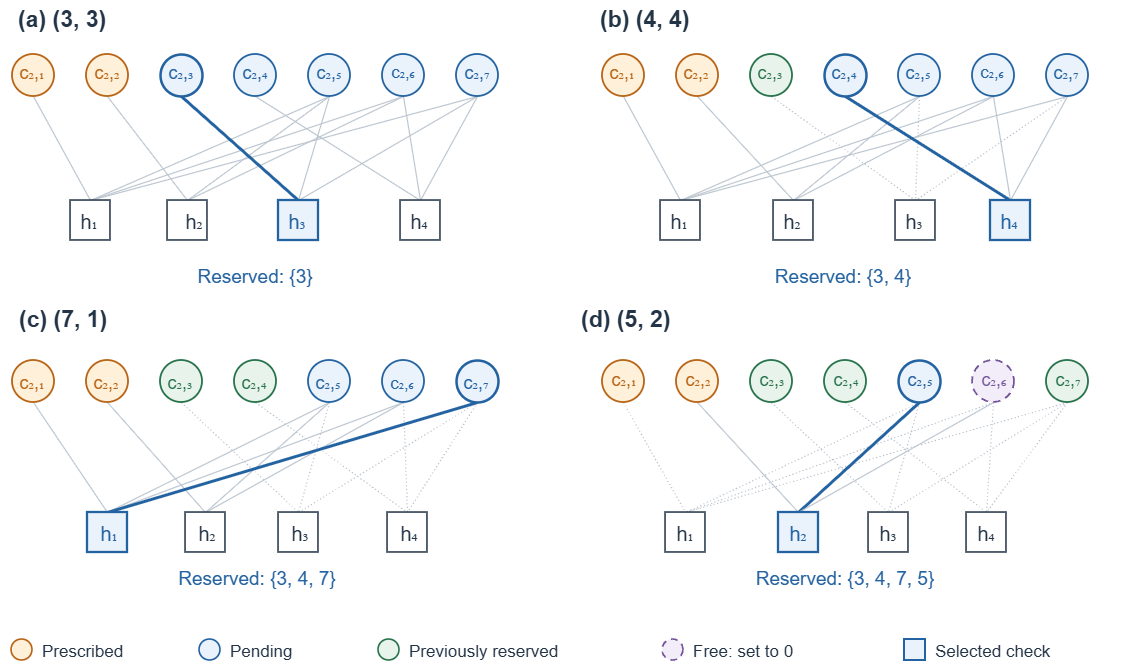}
    \caption{Classical quantization reservation schedule for the running Steane example. Panels (a)--(d) select $(3,3)$, $(4,4)$, $(7,1)$, and $(5,2)$, respectively, removing the selected constraint from further consideration after each reservation. Each pair lists the variable index followed by the constraint index. Blue highlighting identifies the current variable--constraint pair; green nodes indicate previously reserved variables. The prescribed positions are
    $E=\{1,2\}$, while the unreserved variable $c_{2,6}$ remains
    free and is subsequently set to $0$. No binary values are
    evaluated during reservation.}
    \label{fig:steane_quantization}
\end{figure}

\end{example*}

\subsection{Coherent Quantization-Based Cleanup}
\label{subsec:coherent-cleanup}

We now use the quantization schedule from Section~\ref{subsec:c2-quantization} to coherently construct the shift required by \eqref{eq:cleanup-required-condition}, apply it to the auxiliary register, and reverse compute the temporary registers.

Suppose that Algorithm~\ref{alg:erasure-quantization} succeeds. Every constraint has then been reserved. With unreserved variables fixed to zero, the reverse assignments select a completion $c_2(r)$ satisfying~\eqref{eq:codeword-cleanup-condition}. Let $f_E(r)\in\cF_2^{k_2}$ denote its coordinates in the fixed encoder basis, so that $c_2(r)=f_E(r)G_2$. The required shift is therefore $g_E(u)=f_E(uR_E)$. The prepared XOR operations define a linear map on all binary inputs, although a valid completion is required only for targets $r=uR_E$. Let $\widetilde c_2(r)$ denote the full vector of completed graph-variable values, including the $n$ code variables $c_2(r)$ and the $\ell_2$ state variables, if present.

Introduce temporary registers $R$ and $W$ containing $|E|$ and $n+\ell_2$ qubits, respectively. In the following implementation, assume that $G_2$ is systematic in the same encoder basis used in Section~\ref{subsec:inverse-encoding}. Let $W'$ be the $k_2$-qubit subregister of $W$ at these systematic positions; its completed value is then $f_E(r)$. Define the reversible evaluation operations by
\begin{subequations}
\begin{align}
    U_r:\quad \ket{u}_L\ket{r}_R
    &\longmapsto \ket{u}_L\ket{r+uR_E}_R,
    \label{eq:coherent-target-evaluation}\\
    U_q:\quad \ket{r}_R\ket{w}_W
    &\longmapsto \ket{r}_R\ket{w+\widetilde c_2(r)}_W,
    \label{eq:coherent-quantizer-evaluation}\\
    U_a:\quad \ket{x}_X\ket{w'}_{W'}
    &\longmapsto \ket{x+w'}_X\ket{w'}_{W'}.
    \label{eq:coherent-auxiliary-addition}
\end{align}
\end{subequations}
For cleanup, the receiver initializes $R$ and $W$ to $\ket0$ and applies $U_r$, $U_q$, and $U_a$, followed by $U_q^{\dagger}$ and $U_r^{\dagger}$. The addition into $X$ leaves $R$ and $W$ unchanged, so reversing their computation returns both temporary registers to zero. With these temporary registers reset, the resulting operation is
\begin{equation}
\label{eq:coherent-cleanup-action}
    U_{\mathrm{cl}}(E):\quad
    \ket u_L\ket x_X
    \longmapsto
    \ket u_L\ket{x+f_E(uR_E)}_X.
\end{equation}
This realizes \eqref{eq:cleanup-required-action} with $g_E(u)=f_E(uR_E)$.

We now describe how $U_q$ in \eqref{eq:coherent-quantizer-evaluation} executes the assignments specified by $\mathcal Q(E)$ in \eqref{eq:quantizer-reservation-schedule}, in reverse reservation order. Let $w_v$ denote the workspace qubit associated with variable $v$ in $\mathcal G_2$, and let $z_v$ denote its basis value. 
The reason for using $w_v$ instead of $v$ is that the logical state has already been decoded.
After preparing $r=uR_E$, the receiver copies the prescribed values into the corresponding workspace positions:
\begin{equation}
\label{eq:quantizer-prescribed-copy}
    \ket r_R\ket0_{W_E}
    \longmapsto
    \ket r_R\ket r_{W_E}.
\end{equation}
For each pair $(v_\ell,h_\ell)$ in the reservation schedule~\eqref{eq:quantizer-reservation-schedule}, it then applies
\begin{equation}
\label{eq:quantizer-scheduled-assignment}
\begin{aligned}
    &\ket{z_{N(h_\ell)\setminus\{v_\ell\}}}\ket0_{w_{v_\ell}}\\
    &\quad\longmapsto
    \ket{z_{N(h_\ell)\setminus\{v_\ell\}}}
    \ket{\bigoplus_{j\in N(h_\ell)\setminus\{v_\ell\}}z_j}_{w_{v_\ell}},
    \qquad \ell=q,\ldots,1.
\end{aligned}
\end{equation}
Each control value is prescribed, already assigned, or fixed to zero, and each target is still in $\ket0$. Unreserved variables remain zero. When $(v_\ell,h_\ell)$ is reserved, $v_\ell$ has no neighbor among the constraints reserved later. Thus, in reverse reservation order, assigning $v_\ell$ satisfies $h_\ell$ without changing any constraint already satisfied. Since successful reservation includes every constraint, the completed variables satisfy $s_2A_2+c_2B_2=0$. The prescribed coordinates remain equal to $r$, so the code variables in $W$ form a codeword $c_2(r)\in\cC_2$ with $c_{2,E}(r)=r$ and encoder coordinates $f_E(r)$ in $W'$.

The receiver next applies \eqref{eq:coherent-auxiliary-addition}, adding $f_E(r)$ into $X$. It then reverses the assignments in the opposite order:
\begin{equation}
\label{eq:quantizer-scheduled-reverse-computation}
\begin{aligned}
    &\ket{z_{N(h_\ell)\setminus\{v_\ell\}}}
    \ket{\bigoplus_{j\in N(h_\ell)\setminus\{v_\ell\}}z_j}_{w_{v_\ell}}\\
    &\quad\longmapsto
    \ket{z_{N(h_\ell)\setminus\{v_\ell\}}}\ket0_{w_{v_\ell}},
    \qquad \ell=1,\ldots,q.
\end{aligned}
\end{equation}
Finally, reversing \eqref{eq:quantizer-prescribed-copy} clears $W_E$, and applying $U_r^{\dagger}$ clears $R$. Thus, all temporary registers return to zero.

Algorithm~\ref{alg:coherent-cleanup} summarizes the scheduled assignments. It evaluates the reservation schedule in reverse order, adds the resulting encoder coordinates into $X$, and then reverses the computation to clear all temporary registers.

\begin{algorithm}[htbp]
\caption{Coherent Cleanup Using a Quantization Schedule}
\label{alg:coherent-cleanup}
\begin{algorithmic}[1]
\Require Successful schedule $\mathcal Q(E)=((v_1,h_1),\ldots,(v_q,h_q))$; logical register $L$; auxiliary register $X$; temporary registers $R,W$, with $W'\subseteq W$ holding the systematic encoder coordinates
\Ensure $X$ shifted by $f_E(uR_E)$; temporary registers $R,W$ restored to zero

\State Initialize $R$ and $W$ to $\ket0$
\State Apply $U_r$ to prepare $r=uR_E$ in $R$
\Statex \Comment{Begin $U_q$: copy prescribed values and evaluate assignments}
\For{each $i\in E$}
    \State Apply $\operatorname{CNOT}_{R_i\to w_i}$
    \Comment{Copy prescribed values into $W$}
\EndFor

\For{$\ell=q,\ldots,1$}
    \For{$j\in N(h_\ell)\setminus\{v_\ell\}$}
        \State Apply $\operatorname{CNOT}_{w_j\to w_{v_\ell}}$
    \EndFor
\EndFor
\Statex \Comment{End $U_q$: $W$ holds $\widetilde c_2(r)$; $W'$ holds $f_E(r)$}

\For{$j=1,\ldots,k_2$}
    \State Apply $\operatorname{CNOT}_{W'_j\to X_j}$
    \Comment{Add encoder coordinates into $X$}
\EndFor

\Statex \Comment{Begin $U_q^\dagger$: reverse assignments and clear prescribed values}

\For{$\ell=1,\ldots,q$}
    \For{$j\in N(h_\ell)\setminus\{v_\ell\}$}
        \State Apply $\operatorname{CNOT}_{w_j\to w_{v_\ell}}$
    \EndFor
\EndFor

\For{each $i\in E$}
    \State Apply $\operatorname{CNOT}_{R_i\to w_i}$
    \Comment{Clear prescribed positions in $W$}
\EndFor

\Statex \Comment{End $U_q^\dagger$: $W$ is restored to zero}
\State Apply $U_r^{\dagger}$ to clear $R$
\end{algorithmic}
\end{algorithm}

\begin{lemma}[Cleanup complexity]
\label{lem:cleanup-complexity}
Let $\mathcal G_2$ be the constraint graph defined by $s_2A_2+c_2B_2=0$, with $n_v=n+\ell_2$ variable nodes, $n_h=m_2$ check nodes, and $M_2=\operatorname{nnz}(A_2)+\operatorname{nnz}(B_2)$ edges. Suppose reservation succeeds, it produces $\mathcal Q(E)=((v_1,h_1),\ldots,(v_q,h_q))$.

Then, Algorithm~\ref{alg:erasure-quantization} takes $O(n_v+n_h+M_2)$ time and Algorithm~\ref{alg:coherent-cleanup} uses $O(n+M_2+\operatorname{nnz}(R_E))$ CNOT gates. Furthermore, if $A_2$ and $B_2$ have $O(n)$ nonzero entries in total and there are $O(n)$ state variables and checks, Algorithm~\ref{alg:erasure-quantization} takes $O(n)$ time and Algorithm~\ref{alg:coherent-cleanup} requires $O(n+\operatorname{nnz}(R_E))$ CNOT gates, which is $O(n^2)$ in the worst case.
\end{lemma}

\begin{proof}
First consider the time complexity of Algorithm~\ref{alg:erasure-quantization}. Use adjacency lists, constant-time membership flags for $\mathcal U$ and $\mathcal T$, and constant-time queue operations. Initializing the set of unprescribed variables $\mathcal U$ in line~1 takes $O(n_v)$ time. Initializing $\mathcal T$ with all constraint nodes, the empty schedule, and the empty queue in line~2 takes $O(n_h)$ time. Computing the remaining degrees and initializing the queue in lines~3--8 takes $O(n_v+M_2)$ time. Each constraint is removed at most once, so scanning its neighbors and updating the remaining degrees in lines~19--24 traverses at most $M_2$ edges in total. Finding the unique neighbor constraint remaining in line~15 requires scanning the adjacency list of the selected variable; each variable is selected at most once and removed from $\mathcal U$ in line~17, so these scans also take $O(M_2)$ time in total. Each variable enters the queue at most once because its remaining degree only decreases, giving $O(n_v)$ queue operations. Appending reserved pairs and removing reserved constraints take $O(n_h)$ time in total. These observations give the stated $O(n_v+n_h+M_2)$ scheduling bound.

Now consider the gate complexity of Algorithm~\ref{alg:coherent-cleanup}. Preparing the target and reversing its computation in lines~2 and~22 use $2\operatorname{nnz}(R_E)$ CNOT gates. Copying and clearing the prescribed workspace positions in lines~3--5 and~19--21 use $2|E|\leq 2n$ gates. Since $G_2$ is systematic in the chosen encoder basis, adding the coordinates from $W'$ into $X$ in lines~11--13 uses $k_2\leq n$ gates. Each reserved constraint contributes $2(|N(h_\ell)|-1)$ gates for its evaluation and reversal in lines~6--10 and~14--18 and is reserved at most once. Summing these costs over reserved constraints therefore gives at most $2M_2$ gates. These observations give the stated $O(n+M_2+\operatorname{nnz}(R_E))$ gate bound. Under the stated sparsity assumptions, the scheduling time is $O(n)$ and the gate count is $O(n+\operatorname{nnz}(R_E))$. Since $\operatorname{nnz}(R_E)\leq k|E|\leq n^2$, the gate count is $O(n^2)$ in the worst case.
\end{proof}

Applying $U_{\mathrm{cl}}(E)$ to the inverse-encoded branch in~\eqref{eq:post-inverse-branch} gives the factorized state in~\eqref{eq:post-inverse-branch-simplified}, since the implemented shift satisfies~\eqref{eq:cleanup-required-condition}. With all temporary registers returned to zero, we write
\begin{equation}
\label{eq:cleaned-branch}
    \ket{\chi'_b}
    = U_{\mathrm{cl}}(E)\ket{\chi_b}
    = \ket{\psi_L}\otimes\ket{\xi_{E,b}},
\end{equation}
where
\begin{equation}
\label{eq:cleanup-residual-state}
    \ket{\xi_{E,b}}
    =
    \left(
    \frac{1}{\sqrt{2^{k_2}}}
    \sum_{\substack{x'\in\cF_2^{k_2}:\\x'G_{2,E}=b}}
    \ket{x'}_X
    \right)
    \otimes\ket{0}_Z^{\otimes(n-k_1)}\ket{0}_S^{\otimes\ell_1}.
\end{equation}
The residual vector is independent of the logical input, and its squared norm is the branch probability. The same circuit acts on every branch without access to $b$.

\begin{example*}[Steane example: cleanup after inverse encoding.]
For $E=\{1,2\}$, the target preparation \eqref{eq:coherent-target-evaluation} gives $uR_E=(u,u)$. The quantization operation \eqref{eq:coherent-quantizer-evaluation} copies these values into $w_1,w_2$ and executes \eqref{eq:steane-quantizer-schedule} in reverse order,
with all unprescribed workspace variables initially zero. Writing $c_{2,i}$ for the value in $w_i$, the assignments are
\begin{equation}
\label{eq:steane-quantizer-assignments}
\begin{aligned}
    c_{2,5}&=c_{2,2}+c_{2,6}=u,\\
    c_{2,7}&=c_{2,1}+c_{2,5}+c_{2,6}=0,\\
    c_{2,4}&=c_{2,6}+c_{2,7}=0,\\
    c_{2,3}&=c_{2,5}+c_{2,7}=u,
\end{aligned}
\end{equation}
where the unreserved coordinate $c_{2,6}$ remains zero, giving the completed workspace word $(u,u,u,0,u,0,0)$. The systematic positions give $W'=(w_5,w_6,w_7)$, containing
\begin{equation}
\label{eq:steane-cleanup-shift}
    f_E(uR_E)=(c_{2,5},c_{2,6},c_{2,7})=(u,0,0).
\end{equation}
Thus \eqref{eq:coherent-auxiliary-addition} acts as
\begin{equation}
\label{eq:steane-auxiliary-addition}
    \ket{x_1,x_2,x_3}_X\ket{u,0,0}_{W'}
    \longmapsto
    \ket{x_1+u,x_2,x_3}_X\ket{u,0,0}_{W'}.
\end{equation}
Applying $U_q^{-1}$ followed by $U_r^{-1}$ clears the temporary registers. The resulting operation on $L,X$ is
\begin{equation}
\label{eq:steane-cleanup-gate}
    U_{\mathrm{cl}}(E)=\operatorname{CNOT}_{L\to X_1},
\end{equation}
The complete implementation is shown in Fig.~\ref{fig:steane_cleanup}.

\begin{figure}[htbp]
    \centering
    \includegraphics[width=\columnwidth]{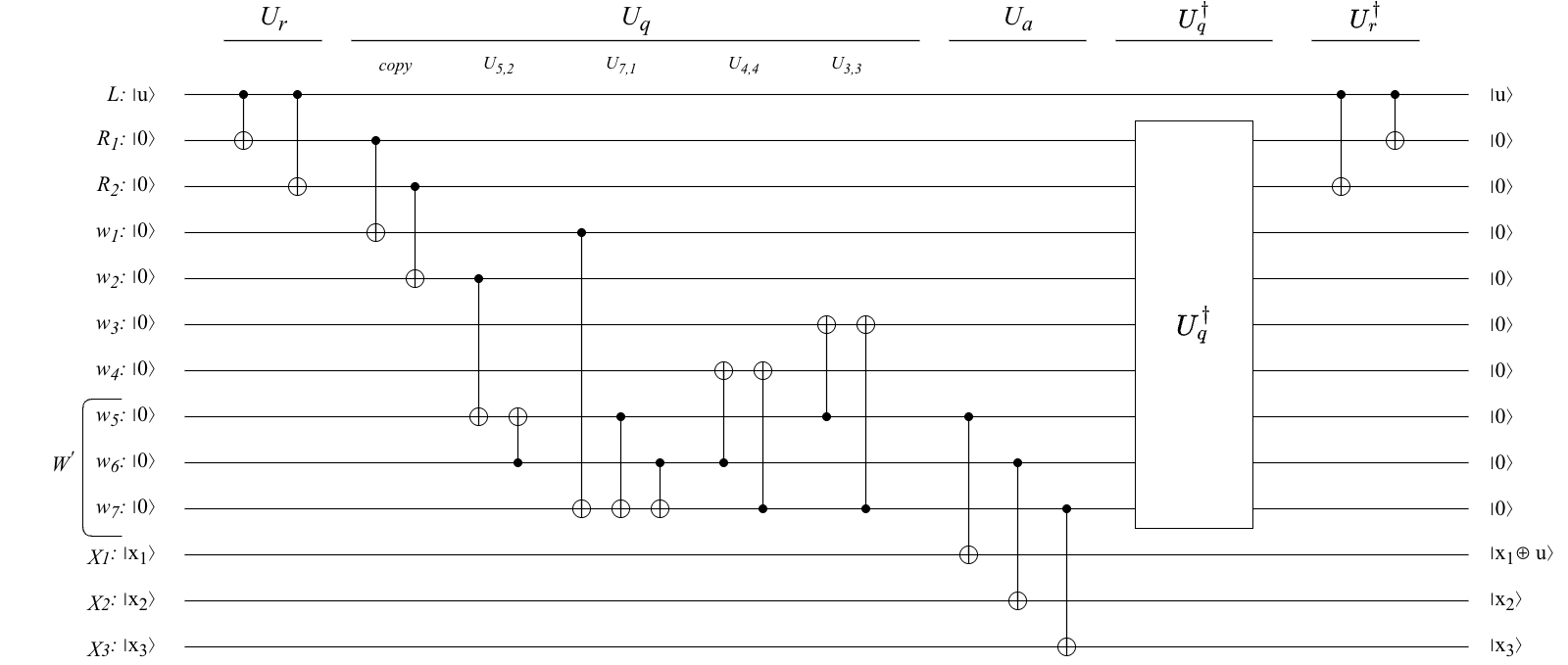}
    \caption{Coherent quantization-based cleanup for the Steane code with $E=\{1,2\}$. The operation $U_r$ prepares $(u,u)$ in $R$, and $U_q$ completes the workspace word by executing the reservation schedule in reverse order. The operation $U_a$ adds the systematic coordinates in $W'=(w_5,w_6,w_7)$ into $X$. Finally, $U_q^{-1}$ and $U_r^{-1}$ return $W$ and $R$ to zero. The net action on $L,X$ is $\operatorname{CNOT}_{L\to X_1}$.}
    \label{fig:steane_cleanup}
\end{figure}

Applying this operation to the four inverse-encoded branches gives
\begin{subequations}
\begin{align}
\label{eq:steane-cleaned-branch}
    \ket{\chi'_{00}}
    &=\ket{\psi_L}\otimes
      \frac{\ket{000}+\ket{110}}{\sqrt8}\otimes\ket{000},\\
\label{eq:steane-cleaned-branch-01}
    \ket{\chi'_{01}}
    &=\ket{\psi_L}\otimes
      \frac{\ket{011}+\ket{101}}{\sqrt8}\otimes\ket{000},\\
\label{eq:steane-cleaned-branch-10}
    \ket{\chi'_{10}}
    &=\ket{\psi_L}\otimes
      \frac{\ket{001}+\ket{111}}{\sqrt8}\otimes\ket{000},\\
\label{eq:steane-cleaned-branch-11}
    \ket{\chi'_{11}}
    &=\ket{\psi_L}\otimes
      \frac{\ket{010}+\ket{100}}{\sqrt8}\otimes\ket{000}.
\end{align}
\end{subequations}
Each branch has squared norm $1/4$ and contains the same recovered logical state $\ket{\psi_L}$.
\end{example*}

\subsection{Correctness of the Proposed Decoder}
\label{subsec:correctness}

We now establish sufficient conditions for exact logical recovery.

\begin{theorem}[Exact Logical Recovery]
\label{prop:exact-logical-recovery}
Consider a pair of nested linear codes
defined by $\cC_2\subseteq\cC_1$ and a fixed erasure pattern $E$. Suppose that:
\begin{enumerate}
    \item Algorithm~\ref{alg:classical-peeling} succeeds on $E$.
    \item Algorithm~\ref{alg:erasure-quantization} succeeds on $E$.
\end{enumerate}
Then coherent reconstruction $U_{rec}(E)$, followed by inverse encoding $U_{\mathrm X}^{\dagger}$ and cleanup $U_{\mathrm{cl}}(E)$, exactly recovers any logical state $\ket{\psi_L}$, with output
\begin{equation}
\label{eq:exact-logical-output}
    \rho_{\mathrm{out},E}=\ket{\psi_L}\!\bra{\psi_L}\otimes\sigma_E,
\end{equation}
where $\sigma_E$ is a normalized density operator independent of the logical input.
\end{theorem}

\begin{proof}
For a pure logical input $\ket{\psi_L}=\sum_u\alpha_u\ket u$, successful classical decoding gives the reconstructed branches in~\eqref{eq:reconstructed-physical-branch}. Inverse encoding produces~\eqref{eq:post-inverse-branch}. Successful reservation guarantees that the reverse assignments produce a valid completion for every prescribed vector, including each required target $uR_E$. Therefore, the shift $g_E(u)=f_E(uR_E)$ satisfies~\eqref{eq:cleanup-required-condition}. Cleanup gives the factorization in~\eqref{eq:cleaned-branch}, with all temporary registers returned to zero.

Summing over the inaccessible branches yields
\begin{equation}
\label{eq:output-factorization}
\begin{aligned}
    \rho_{\mathrm{out},E}
    &=\sum_b\ket{\chi'_b}\!\bra{\chi'_b} = 
    \ket{\psi_L}\!\bra{\psi_L}\otimes\sigma_E,
\end{aligned}
\end{equation}
where
\begin{equation}
\label{eq:residual-density}
    \sigma_E
    =\sum_{b\in\cF_2^{|E|}}
    \ket{\xi_{E,b}}\!\bra{\xi_{E,b}},
\end{equation}
and the residual vectors $\ket{\xi_{E,b}}$ in~\eqref{eq:cleanup-residual-state} are independent of the logical input. Moreover,
\begin{equation}
\label{eq:residual-normalization}
\begin{aligned}
    \operatorname{Tr}(\sigma_E)
    &=\frac{1}{2^{k_2}}
    \sum_{b\in\cF_2^{|E|}}
    \big|\{x\in\cF_2^{k_2}:xG_{2,E}=b\}\big|
    &=1,
\end{aligned}
\end{equation}
since each of the $2^{k_2}$ vectors $x$ contributes to exactly one term, namely $b=xG_{2,E}$. 
\end{proof}

These conditions are sufficient for the proposed decoder. Note that failure of a particular classical decoding or quantization algorithm does not imply that the erasure pattern is uncorrectable by every recovery method.

We conclude this section by noting that the overall classical decoder time complexity is $O(n)$ and the overall gate complexity is $O(n^2)$. Furthermore, the $O(n^2)$ complexity manifests itself whenever a generator matrix ($R$ or $G_2$) is involved. One potential research direction is to incorporate known results from LDPC code encoding~\cite [Appendix~A]{Richardson_Urbanke_2008} to further reduce the gate complexity to $O(n)$.

\section{Architecture II: Fully Quantum Belief Propagation with Coherent Reconstruction}
\label{sec:quantum-bp-architecture}

We now consider a quantum implementation of belief propagation (BP) for erasure decoding of a pair of nested linear codes.
 
Unlike Architecture I, where we compute the peeling schedule classically and then translate it into coherent reconstruction gates, the present architecture performs BP computations directly on quantum registers. The decoder employs reversible ternary operations to evaluate parity-check constraints, identify recoverable variables, and reconstruct their values without measuring the surviving code-variable qubits. This coherent reconstruction is the first stage of quantum state healing, restoring erased code variables and recovering the state variables while preserving coherence within each erasure-environment branch. Each iteration consists of the QBP computation, the reversible gate $U_{\mathrm{swap}}^{\otimes n_v}$, and ancilla reset. The latter two operations together constitute the swap-and-reset (S\&R) procedure. We reuse the same ancilla registers throughout a fixed iteration budget, as described below. Once reconstruction is complete, the receiver applies the inverse binary encoder and coherent quantization-based cleanup described in Architecture I. The cleanup uses a classically prepared reservation schedule and a temporary quantum workspace that is returned to zero. Thus, the distinction between the two architectures lies primarily in the implementation of the code variable reconstruction stage.

\begin{figure}[htbp]
    \centering
    \includegraphics[width=\columnwidth]{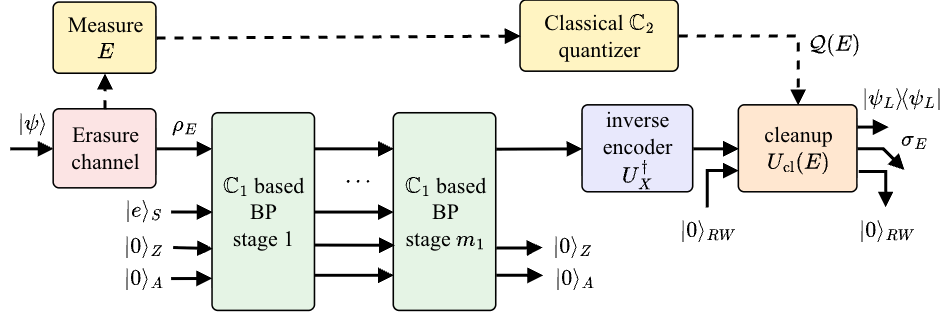}
    \caption{Proposed quantum erasure decoding architecture II.}
    \label{fig:architecture2}
\end{figure}

\subsection{Quantum Registers and Decoder Initialization}
\label{subsec:qbp-registers}

For the QBP construction, we use the constraint graph $\mathcal G_1$ defined by $s(c)A_1+cB_1=0$, with $n_v=n+\ell_1$ variable nodes and $m_1$ check nodes. The variable registers consist of the $n$ code-variable qutrits $Y$ and the $\ell_1$ state-variable qutrits $S$, each with basis $\{\ket0,\ket1,\ket e\}$, where $e\equiv2$ represents an erasure. We denote the combined variable register by $V=(Y,S)$. The check register $Z$ contains $m_1$ qutrits used to evaluate local parity constraints, and the update register $A$ contains $n_v$ qutrits, one for each code variable or state variable.

Initialize the check and update registers to zero. Each qutrit in $S$ is initialized in $\ket0$ and then mapped to $\ket e$ by a unitary fixed single-qutrit, marking its value as unknown to BP. The received $Y$ remain unchanged, including those in the erasure state $\ket e$. Using $S$ to denote the set of state nodes as well, QBP therefore starts with unresolved variables $E\cup S$ and applies the same update rules to code variables and state variables. If BP resolves this enlarged set, it produces $\ket c_Y\ket {s(c)}_S$, satisfying $s(c)A_1+cB_1=0$. The reconstruction stage uses $\ell_1+m_1+n_v$ auxiliary qutrits in addition to the $n$ received code-variable qutrits. The check and update registers are recycled between iterations. This count concerns reconstruction; the subsequent coherent cleanup has the workspace requirements described in Section~\ref{subsec:coherent-cleanup}.

Unlike Architecture I, reconstruction of the code variables and state variables does not require a classically prepared peeling schedule. Ternary quantum operations determine which variables can be recovered at each iteration. After successful reconstruction, the joint inverse encoder returns the state register $S$ to zero. We still prepare a quantization reservation schedule classically for the final cleanup.

\subsection{Reversible Quantum BP Iteration}
\label{subsec:quantum-bp-iteration}

We first describe the classical ternary update rules that the QBP circuit implements reversibly. Here $y$ denotes the vector of values in $V$, so the check-node and variable-node rules apply to both code variables and state variables.

For each check node $c$, define the ternary-valued function
\begin{equation}
\label{eq:qbp-check-update}
g_c(y_{N(c)})=
\begin{cases}
\displaystyle
\bigoplus_{\substack{i\in N(c)\\y_i\neq e}}y_i,
& \bigl|\{i\in N(c):y_i=e\}\bigr|=1,\\[6pt]
e,
& \bigl|\{i\in N(c):y_i=e\}\bigr|>1,\\
0,
& \bigl|\{i\in N(c):y_i=e\}\bigr|=0.
\end{cases}
\end{equation}
The conditions in~\eqref{eq:qbp-check-update} are evaluated directly from the input symbols, so the gate does not require advance classical knowledge of the unresolved erasure locations. The first case corresponds to a degree-one check: the value of its unique erased neighbor can be determined from the remaining variables. A check containing multiple erased variables returns $e$, indicating that the corresponding parity constraint cannot yet resolve an erased variable.

The variable-node refinement function is defined as
\begin{equation}
\label{eq:qbp-variable-update}
h_v(y_v,z_{N(v)})=
\begin{cases}
y_v, & y_v\neq e,\\
z_{c_v(z)}, &  y_v=e,\ \exists c\in N(v):z_c\neq e, \\
e, & \text{otherwise},
\end{cases}
\end{equation}
where $c_v(z)$ denotes the first check satisfying $z_c\neq e$ in a fixed ordering of $N(v)$. This convention makes $h_v$ well-defined on every computational-basis input. For inputs arising from a valid codeword, all non-erased check values neighboring an erased variable agree on its recovered value.

For the register operations below, we identify the qutrit labels $0,1,e$ with the integers $0,1,2$, respectively. Addition modulo three therefore acts on these labels, whereas the parity sums in $g_c$ remain binary XORs. For a target initialized to $\ket0$, this addition writes the computed ternary value into the target. For an arbitrary target value, it is an invertible cyclic shift, whose inverse is subtraction modulo three. We implement the check-node computation using the reversible transformation
\begin{equation}
\label{eq:qbp-check-gate}
U_g:
\ket y_V\ket z_Z
\longmapsto
\ket y_V
\ket{z+g(y)\bmod3}_Z.
\end{equation}
where $g(y)$ contains the values computed by all check nodes, and addition is performed component-wise modulo three.

Similarly, the variable-node update is implemented by
\begin{equation}
\label{eq:qbp-update-gate}
U_h:\quad
\ket y_V\ket z_Z\ket a_A
\longmapsto
\ket y_V\ket z_Z
\ket{a+h(y,z)\bmod3}_A.
\end{equation}

Both transformations are reversible because the original variable and check registers are preserved, and the target-register additions are invertible. One QBP iteration first computes the check-node values, then writes the variable-node refinements into the update ancillas, and finally reverses the check-node computation. The code-variable and state-variable qutrits in $V$ remain unchanged during this operation. 
\begin{equation}
\label{eq:qbp-composition}
U_{\mathrm{QBP}}
=
U_g^\dagger U_h U_g.
\end{equation}

Consequently,
\begin{equation}
\label{eq:qbp-action}
U_{\mathrm{QBP}}
\ket y_V\ket0_Z^{\otimes m_1}\ket0_A^{\otimes n_v}
=
\ket y_V\ket0_Z^{\otimes m_1}
\ket{h(y,g(y))}_A.
\end{equation}

Since the transformations are unitary permutations of computational-basis states, the same circuit acts coherently on arbitrary superpositions of received codewords.

\subsection{Swap and Reset (S\&R)}
\label{subsec:S-R}
Following the QBP computation, each register $V_v$ is paired with its update ancilla $A_v$. The swap-and-reset (S\&R) procedure consists of a reversible two-qutrit gate $U_{\mathrm{swap}}$, followed by reset of the ancilla to $\ket0$ for reuse. Here, ``swap'' denotes the transfer of the updated value into the variable register according to the mappings below.

For a previously known variable, the QBP update satisfies $a_v=y_v$; for an erased variable, $a_v$ is either its recovered binary value or the erasure symbol $e$. On these inputs, the reversible gate acts as
\begin{subequations}
\label{eq:S-R-required}
\begin{align}
\ket b_V\ket b_A &\longmapsto \ket b_V\ket0_A, \qquad b\in\cF_2,\\
\ket e_V\ket b_A &\longmapsto \ket b_V\ket e_A, \qquad b\in\cF_2,\\
\ket e_V\ket e_A &\longmapsto \ket e_V\ket0_A.
\end{align}
\end{subequations}
These transformations preserve known variables, reconstruct recoverable erasures, and retain unresolved erasures. They leave each update ancilla in $\ket e$ if its variable was newly recovered and in $\ket0$ otherwise. Table~\ref{tab:S-R} extends these mappings to a permutation of all nine computational-basis states, thereby defining a unitary operator $U_{\mathrm{swap}}$. The unitary performs the transfer step; it does not by itself reset every update ancilla to $\ket0$.

\begin{table}[htbp]
\centering
\begin{tabular}{cc|cc}
\hline
$y_v$ & $a_v$ & $\tilde y_v$ & $\tilde a_v$\\
\hline
0 & 0 & 0 & 0\\
0 & 1 & 0 & 1\\
0 & $e$ & 1 & 1\\
1 & 0 & $e$ & 1\\
1 & 1 & 1 & 0\\
1 & $e$ & $e$ & $e$\\
$e$ & 0 & 0 & $e$\\
$e$ & 1 & 1 & $e$\\
$e$ & $e$ & $e$ & 0\\
\hline
\end{tabular}
\caption{The reversible two-qutrit gate $U_{\mathrm{swap}}$ used in the swap-and-reset procedure, with $U_{\mathrm{swap}}\ket{y_v}\ket{a_v}=\ket{\widetilde y_v}\ket{\widetilde a_v}$. The four inputs not encountered during ideal QBP evolution complete the permutation. The subsequent reset returns the ancilla to $\ket0$.}
\label{tab:S-R}
\end{table}

The reversible gate acts independently on each variable--ancilla pair:
\begin{equation}
\label{eq:qbp-swap-parallel}
U_{\mathrm{swap}}^{\otimes n_v}
=
\bigotimes_{v=1}^{n_v}U_{\mathrm{swap}}^{(v)}.
\end{equation}
Here $U_{\mathrm{swap}}^{(v)}$ acts on $V_v,A_v$. The check register $Z$, which has already been returned to zero by $U_g^\dagger$, is unaffected.

We treat reset as a primitive quantum channel that returns an ancilla to $\ket0$, irrespective of its previous state. Its physical implementation is not specified here. The factorization below establishes that resetting the ancillas preserves the reconstructed quantum information. To justify this reset, fix an erasure pattern $E$ and initialize the state variables as unknown, giving the enlarged unresolved set $E\cup S$. Let $\ket{\Phi_{E,b}^{(\ell)}}_V$ denote the unnormalized joint state register in erasure-environment branch $b$ before iteration $\ell$, and let $\eta^{(\ell)}(E\cup S)\in\{0,e\}^{n_v}$ have entry $e$ precisely at the variable locations newly recovered in that iteration. By~\eqref{eq:qbp-check-update} and~\eqref{eq:qbp-variable-update}, these locations depend only on the unresolved variable locations and the graph, not on the codeword values or the branch label $b$. Hence, \eqref{eq:qbp-action} and~\eqref{eq:S-R-required} give
\begin{equation}
\label{eq:qbp-ancilla-factorization}
\begin{aligned}
&U_{\mathrm{swap}}^{\otimes n_v}U_{\mathrm{QBP}}
\left(
\ket{\Phi_{E,b}^{(\ell)}}_V
\ket0_Z^{\otimes m_1}
\ket0_A^{\otimes n_v}
\right)\\
&\qquad=
\ket{\Phi_{E,b}^{(\ell+1)}}_V
\ket0_Z^{\otimes m_1}
\ket{\eta^{(\ell)}(E\cup S)}_A.
\end{aligned}
\end{equation}
The ancilla state is independent of the codeword values and the branch label $b$, so it factors from the joint variable-register state and records only which positions were recovered in that iteration. Resetting $A$ to $\ket0^{\otimes n_v}$ therefore completes S\&R without disturbing the joint state of $Y,S$. If BP resolves every variable in $E\cup S$, each computational-basis input $\ket e_E^{\otimes|E|}\ket{c_{\bar E}}_{\bar E}\ket e_S^{\otimes\ell_1}$ is reconstructed as $\ket c_Y\ket s_S$, where $s(c)A_1+cB_1=0$. By linearity and~\eqref{eq:qbp-ancilla-factorization}, successful QBP produces the same joint reconstructed state as Architecture~I, given by~\eqref{eq:reconstructed-physical-density}--\eqref{eq:reconstructed-physical-branch}.

For the constraint graph with $m_1$ check nodes, any enlarged erasure pattern $E\cup S$ that BP successfully resolves is reconstructed in at most $m_1$ QBP--S\&R iterations. Indeed, each productive iteration contains a check with exactly one unresolved neighbor, and that check has no unresolved neighbors after the iteration, so it cannot be responsible for recovery in a later iteration. We may therefore use a fixed budget of $m_1$ iterations, with zero-initialized update ancillas at the start of each iteration. If reconstruction finishes earlier, the remaining iterations leave the joint state of $Y,S$ unchanged. If BP stalls with unresolved variables, further identical iterations do not resolve them.

We reuse the same check register $Z$ and update register $A$ throughout reconstruction. In each iteration, $U_g^\dagger$ returns $Z$ to $\ket0^{\otimes m_1}$, and the reset following $U_{\mathrm{swap}}^{\otimes n_v}$ returns $A$ to $\ket0^{\otimes n_v}$. Both registers are therefore ready for reuse in the next iteration. Reset is required before reusing $A$ because $U_h$ adds the new update to its existing contents rather than overwriting them. This construction uses $m_1$ check qutrits and $n_v$ update qutrits, in addition to the $n$ code-variable and $\ell_1$ state-variable qutrits, independent of the number of iterations.

If no unresolved variables remain, the QBP update reproduces the existing variable values, and $U_{\mathrm{swap}}$ clears the update ancillas while preserving the combined variable register. Thus, for every state $\ket\psi_V$ supported on the binary subspace,
\begin{equation}
\label{eq:qbp-no-erasure}
\begin{aligned}
&U_{\mathrm{swap}}^{\otimes n_v}U_{\mathrm{QBP}}
\left(
\ket\psi_V
\ket0_Z^{\otimes m_1}
\ket0_A^{\otimes n_v}
\right)\\
&\qquad=
\ket\psi_V
\ket0_Z^{\otimes m_1}
\ket0_A^{\otimes n_v}.
\end{aligned}
\end{equation}
The subsequent reset leaves these zero-valued ancillas unchanged. Thus, additional iterations leave a fully reconstructed joint state unperturbed. Even when the channel introduces no physical erasures, the state variables are initially unresolved and must still be recovered. Joint inverse encoding and the cleanup procedure remain necessary to complete logical recovery after reconstruction.

\begin{example*}[Steane example: quantum BP reconstruction.]

We illustrate the reconstruction of one computational-basis term in one of the erasure-environment branches given in~\eqref{eq:steane-erased-branches}. For $E=\{1,2\}$, let $c\in\cC_1$ be a codeword contributing to the chosen branch, so that $c_E=b$. The corresponding initial variable-register state is $\ket{y^{(0)}}_Y$, where
\begin{equation}
    y^{(0)}=(e,e,c_3,c_4,c_5,c_6,c_7).
\end{equation}
We use a single ancilla register $A$, which is reset between iterations.

In the first iteration, checks 1 and 2 each contain two erased neighbors, while check 3 has only one. Thus \eqref{eq:qbp-check-update} gives
\begin{equation}
    g(y^{(0)})=(e,e,c_3\oplus c_4\oplus c_7)=(e,e,c_1).
\end{equation}
The variable update recovers position 1, giving
\begin{equation}
    h(y^{(0)},g(y^{(0)}))=(c_1,e,c_3,c_4,c_5,c_6,c_7)=y^{(1)}.
\end{equation}
The operation $U_g$ writes the check values into $Z$, $U_h$ writes the variable updates into $A$, and $U_g^\dagger$ returns $Z$ to zero. The subsequent gate $U_{\mathrm{swap}}^{\otimes7}$ transfers the recovered value into $Y$, leaving the ancilla register $A$ in $\ket{e000000}$ before reset.
\begin{equation}
\label{eq:steane-qbp-first-iteration}
\begin{aligned}
    &\ket{y^{(0)}}_Y\ket{000}_Z\ket{0000000}_A\\
    &\quad\xrightarrow{U_g}
    \ket{y^{(0)}}_Y\ket{e,e,c_1}_Z\ket{0000000}_A\\
    &\quad\xrightarrow{U_h}
    \ket{y^{(0)}}_Y\ket{e,e,c_1}_Z\ket{y^{(1)}}_A\\
    &\quad\xrightarrow{U_g^\dagger}
    \ket{y^{(0)}}_Y\ket{000}_Z\ket{y^{(1)}}_A\\
    &\quad\xrightarrow{U_{\mathrm{swap}}^{\otimes7}}
    \ket{y^{(1)}}_Y\ket{000}_Z\ket{e000000}_A.
\end{aligned}
\end{equation}
Figure~\ref{fig:steane-qbp-circuit} shows the complete first QBP--S\&R iteration, including ancilla reset.

\begin{figure*}[t]
    \centering
    \includegraphics[width=\textwidth]{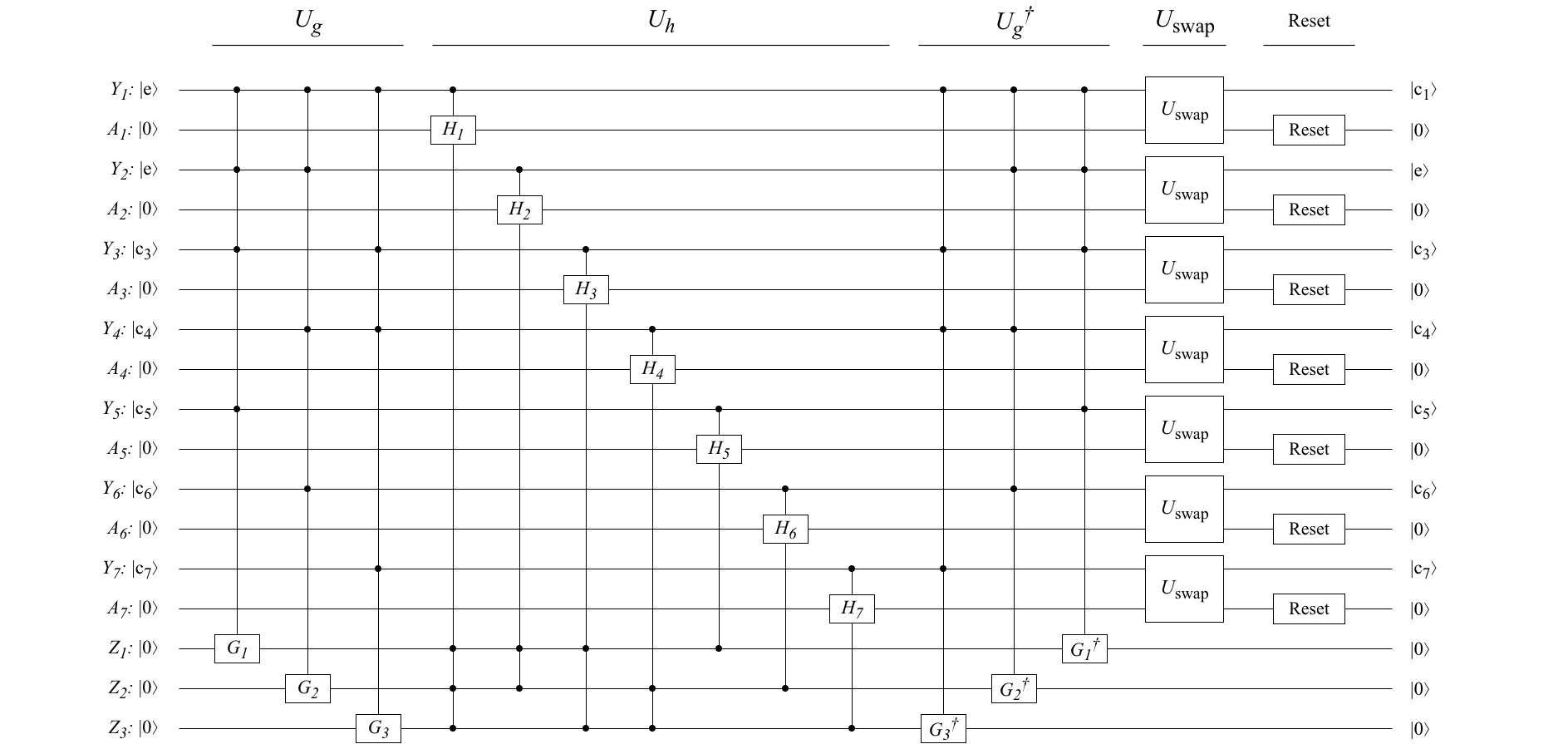}
    \caption{First QBP--S\&R iteration for the Steane example with $E=\{1,2\}$. The gates $G_c$ compute check values, $H_v$ write updated values into the ancilla register $A$, and $G_c^\dagger$ clear $Z$. Filled dots indicate inputs to the ternary functions, not binary controls. The gates $U_{\mathrm{swap}}$ recover position 1, while position 2 remains erased. Each block labeled Reset returns the corresponding ancilla to $\ket0$ for reuse. Both $A$ and $Z$ finish in zero. The displayed states follow one computational-basis term within an erasure-environment branch.}
    \label{fig:steane-qbp-circuit}
\end{figure*}

After $U_{\mathrm{swap}}^{\otimes7}$ and before reset, the ancilla state $\ket{e000000}_A$ records only that position 1 was recovered; it is independent of the codeword values and the branch label. Resetting $A$ returns it to $\ket{0000000}_A$ without disturbing the physical state. This completes the first S\&R procedure and prepares the same ancilla register for reuse in the second iteration.

Since all check updates use the variable values at the start of the iteration, the recovered value $c_1$ becomes available to checks 1 and 2 only in the second iteration. Their outputs are now
\begin{equation}
\begin{aligned}
    g(y^{(1)})
    &=(c_1\oplus c_3\oplus c_5,\ c_1\oplus c_4\oplus c_6,\ 0)\\
    &=(c_2,c_2,0).
\end{aligned}
\end{equation}
Both checks resolve position 2, while check 3 returns zero because it has no erased neighbors. Thus $h(y^{(1)},g(y^{(1)}))=c$, and the second iteration acts as
\begin{equation}
\label{eq:steane-qbp-second-iteration}
\begin{aligned}
    &\ket{y^{(1)}}_Y\ket{000}_Z\ket{0000000}_A\\
    &\quad\xrightarrow{U_g}
    \ket{y^{(1)}}_Y\ket{c_2,c_2,0}_Z\ket{0000000}_A\\
    &\quad\xrightarrow{U_h}
    \ket{y^{(1)}}_Y\ket{c_2,c_2,0}_Z\ket c_A\\
    &\quad\xrightarrow{U_g^\dagger}
    \ket{y^{(1)}}_Y\ket{000}_Z\ket c_A\\
    &\quad\xrightarrow{U_{\mathrm{swap}}^{\otimes7}}
    \ket c_Y\ket{000}_Z\ket{0e00000}_A.
\end{aligned}
\end{equation}
Reset returns the same ancilla register $A$ to $\ket{0000000}_A$, completing the second S\&R procedure. In both iterations, the update-ancilla state before reset depends only on the recovered location, not on the codeword values or the erasure-environment branch. Consequently, the same two iterations coherently reconstruct the superposition within every branch, as established by~\eqref{eq:qbp-ancilla-factorization}. Both $A$ and $Z$ are reused throughout; no fresh ancilla register is introduced.

This reconstruction provides an example of what we informally call ``quantum state healing.'' The reconstruction finishes after two iterations. A third iteration, if included to meet the fixed budget $n-k_1=3$, leaves the reconstructed state unchanged. More generally, when no erasures are present, a complete QBP--S\&R iteration with zero-initialized ancillas acts as
\begin{equation}
    \ket c_Y\ket{000}_Z\ket{0000000}_A
    \longmapsto
    \ket c_Y\ket{000}_Z\ket{0000000}_A.
\end{equation}
By linearity, the complete iteration also leaves any superposition of these codewords unchanged.
\end{example*}

\subsection{Inverse Binary Encoding and Coherent Quantization-Based Cleanup}
\label{subsec:qbp-final-cleanup}

Successful QBP reconstruction produces the same joint branches on $Y, S$ as Architecture~I. The receiver therefore applies the joint inverse binary encoder of Section~\ref{subsec:inverse-encoding}, which exposes the logical and auxiliary registers and returns $S$ to zero, followed by the coherent cleanup operation of Section~\ref{subsec:coherent-cleanup}. Under the quantization condition of Theorem~\ref{prop:exact-logical-recovery}, these stages give the output in~\eqref{eq:exact-logical-output}, with a residual state independent of the logical input.

\section{Conclusions}
\label{sec:conclusions}

In this work, we presented a coherent erasure-recovery framework for nested binary linear codes that integrates reconstruction, inverse binary encoding, and quantization-based cleanup. 
We adopt a perspective motivated by the study of quantum capacity of quantum channels, namely that a good channel code (packing code), and a good source code (covering code) are required for capacity achievability, and we build a decoder that operates directly on these two codes.
A key insight is that merely reconstructing valid codewords within the erasure-environment branches is not enough for complete logical recovery. After performing inverse encoding, the encoder coordinates of an inner-code completion define an auxiliary shift that eliminates the logical-label dependence of the residual state. When reconstruction is successful, and the quantizer provides valid completions for all required targets, the receiver can accurately recover the logical state without needing access to the erased values or the environment branch labels.

The two presented architectures implement this recovery framework through different reconstruction mechanisms. Architecture~I establishes peeling decoding and peeling quantization schedules using classical decoders, then performs the corresponding binary operations coherently. In contrast, Architecture~II directly evaluates the reconstruction updates on qutrit registers and subsequently applies the same inverse encoding and cleanup operations. The ancillary states used for updates can be reset without disrupting the logical information, as they depend solely on the erasure pattern and the status of recovery.
The Steane example illustrates both reconstruction mechanisms, as well as the auxiliary shift necessary for completing recovery. 

Looking ahead, future work will focus on implementing the cleanup operation using ternary gates without relying on any classical measurement. Additional implementation questions that need addressing include the cost of inverse encoding and target preparation, the implementation of ternary gates, and the trade-offs among circuit depth, workspace, and ancilla reset.

\bibliographystyle{IEEEtran}
\bibliography{achilleas18abrv,references}

\end{document}